\documentclass[twocolumn]{aastex63}

\usepackage[]{natbib}
\usepackage[]{graphicx}
\usepackage[]{amsmath, amssymb}

\usepackage[outdir=./figs/]{epstopdf}
\usepackage{array}
\usepackage{booktabs}
\usepackage{xspace}
\usepackage{float}
\usepackage[caption=false]{subfig}
\usepackage{color}

\newcommand{\Ha}{H$\alpha$\xspace}
\newcommand{\Hb}{H$\beta$\xspace}
\newcommand{\OII}{[\ion{O}{2}]\xspace}
\newcommand{\OIII}{[\ion{O}{3}]\xspace}
\def\ecs{{ergs~cm$^{-2}$~s$^{-1}$}}
\newcommand{\Msun}{M$_\odot$}

\newcommand{\nsample}{1951} 
\newcommand{\threedhst}{\mbox{3D-HST}}
\newcommand{\nsamplefull}{1951} 
\newcommand{\nsampleunmasked}{1035} 
\newcommand{\nsamplefluxcutfitmasked}{859} 
\newcommand{\nsimsample}{100}
\newcommand{\nagnfull}{44}
\newcommand{\paperone}{B19} 

\newcommand{\ie}{i.e.,\ }

\newcommand{\HST}{{\sl HST}}

\newcommand{\RST}{{\sl Roman}\xspace}

\def\lesssim{\mathrel{\hbox{\rlap{\hbox{\lower4pt\hbox{$\sim$}}}\hbox{$<$}}}}

\begin{document}

\title{The $z \sim 2$ \OIII Luminosity Function of Grism-selected Emission-line Galaxies}

\shorttitle{The $z \sim 2$ \OIII Luminosity Function}

\correspondingauthor{William P. Bowman}
\email{wpb.astro@gmail.com}

\author[0000-0003-4381-5245]{William P. Bowman}
\affiliation{Department of Astronomy \& Astrophysics, The Pennsylvania
State University, University Park, PA 16802}
\affiliation{Institute for Gravitation and the Cosmos, The Pennsylvania State University, University Park, PA 16802}

\author[0000-0002-1328-0211]{Robin Ciardullo}
\affiliation{Department of Astronomy \& Astrophysics, The Pennsylvania
State University, University Park, PA 16802}
\affiliation{Institute for Gravitation and the Cosmos, The Pennsylvania State University, University Park, PA 16802}

\author[0000-0003-2307-0629]{Gregory R. Zeimann}
\affiliation{Hobby Eberly Telescope, University of Texas, Austin, Austin, TX, 78712, USA}

\author[0000-0001-6842-2371]{Caryl Gronwall}
\affiliation{Department of Astronomy \& Astrophysics, The Pennsylvania
State University, University Park, PA 16802}
\affiliation{Institute for Gravitation and the Cosmos, The Pennsylvania State University, University Park, PA 16802}

\author[0000-0002-8434-979X]{Donghui Jeong}
\affiliation{Department of Astronomy \& Astrophysics, The Pennsylvania
State University, University Park, PA 16802}
\affiliation{Institute for Gravitation and the Cosmos, The Pennsylvania State University, University Park, PA 16802}

\author{Gautam Nagaraj}
\affiliation{Department of Astronomy \& Astrophysics, The Pennsylvania
State University, University Park, PA 16802}
\affiliation{Institute for Gravitation and the Cosmos, The Pennsylvania State University, University Park, PA 16802}

\author{Cullen Abelson}
\affiliation{Department of Astronomy \& Astrophysics, The Pennsylvania
State University, University Park, PA 16802}

\author[0000-0002-4974-1243]{Laurel H. Weiss}
\affiliation{Department of Astronomy \& Astrophysics, The Pennsylvania
State University, University Park, PA 16802}

\author[0000-0001-8440-3613]{Mallory Molina}
\affiliation{eXtreme Gravity Institute, Department of Physics, Montana State University, Bozeman, MT 59717, USA}

\author{Donald P. Schneider}
\affiliation{Department of Astronomy \& Astrophysics, The Pennsylvania
State University, University Park, PA 16802}
\affiliation{Institute for Gravitation and the Cosmos, The Pennsylvania State University, University Park, PA 16802}

\begin{abstract}
\noindent
Upcoming missions such as {\sl Euclid} and the {\sl Nancy Grace Roman Space Telescope} (\RST) will use emission-line selected galaxies to address a variety of questions in cosmology and galaxy evolution in the $z>1$ universe. The optimal observing strategy for these programs relies upon knowing the number of galaxies that will be found and the bias of the galaxy population.  Here we measure the \OIII~$\lambda 5007$ luminosity function for a vetted sample of \nsamplefull\ $m_{\rm J+JH+H} < 26$ galaxies with unambiguous redshifts between $1.90 < z < 2.35$, which were selected using \HST/WFC3 G141 grism frames made available by the 3D-{\sl Hubble Space Telescope} program. These systems are directly analogous to the galaxies that will be identified by the {\sl Euclid} and \RST\ missions, which will utilize grism spectroscopy to find \OIII~$\lambda 5007$-emitting galaxies at $0.8 \lesssim z \lesssim 2.7$ and $1.7 \lesssim z \lesssim 2.8$, respectively.  We interpret our results in the context of the expected number counts for these upcoming missions.
Finally, we combine our dust-corrected \OIII\ luminosities with rest-frame ultraviolet star formation rates to present 
a new calibration
of the SFR density associated with $1.90 < z < 2.35$ \OIII-emitting galaxies.  We find that these grism-selected galaxies contain roughly half of the total star formation activity at $z\sim2$.
\end{abstract}

\keywords{galaxies: evolution -- galaxies: high-redshift -- cosmology: observations}

\section{Introduction}
\label{sec:intro}

The $\Lambda$CDM cosmological model has proven successful at explaining a wide variety of astrophysical phenomena and has withstood stringent tests of its validity (see, e.g., the reviews by \citet{bull2016} and \citet{amendola2018} and references therein).  While $\Lambda$CDM has facilitated important breakthroughs in our understanding of the origin and evolution of the universe, so too has it introduced some of the biggest mysteries, including the nature of dark energy and its effect on the acceleration of the universe \citep{riess1998}.
Ongoing and future survey missions such as the Dark Energy Spectroscopic Survey \citep{DESI}, the Hobby-Eberly Telescope Dark Energy Experiment (HETDEX; Hill et al.~2021 submitted, Gebhardt et al.~2021, submitted), {\sl Euclid} \citep{laureijs2011, laureijs2012} and the {\sl Nancy Grace Roman Space Telescope} \citep[\RST;][]{green2012, spergel2015} will explore dark energy by mapping out the three-dimensional positions of galaxies over vast, contiguous volumes of cosmic spacetime using emission lines in the rest-frame ultraviolet and optical. The achievable precision of these experiments is directly proportional to the number of galaxies detected by the surveys. Accurate galaxy emission-line luminosity functions are thus critically important for planning the optimal observing strategies for these missions, as they directly inform the number of tracers one expects to find in a given exposure time.

The {\sl Euclid} and \RST satellites will have infrared grism spectrographs capable of measuring the wavelengths and total fluxes of emission-lines between about $\sim 1.0\,\mu$m and $\sim 2.0\,\mu$m.  In addition to mapping out large scale structure, these data will enable uniform measures of galactic star-formation rates and internal extinctions for hundreds of thousands of galaxies via their Balmer emission lines.  However, beyond $z \gtrsim 2$, \Ha shifts out of the grisms' bandpasses, leaving \OIII as the dominant feature.  At these redshifts, the missions will need to rely on forbidden \OIII~$\lambda 5007$ to trace the galaxy populations of the universe.

Existing measurements of the \OIII luminosity function have yielded important insights into the evolution of the \OIII~$\lambda 5007$ line across cosmic time \citep{ly2007, pirzkal2013, colbert2013, drake2013, sobral2015, khostovan2015, mehta2015, comparat2016, khostovan2020, hayashi2020}. However, most of these studies have either been focused at low ($z < 1$) redshift, have high equivalent width thresholds, survey small volumes of the high-$z$ universe, or are defined by color-selection, rather than through the detection of emission lines.  The one notable exception is the study by \citet{mehta2015}, who, building upon the work of \citet{colbert2013}, identified \OIII emitters out to $z \sim 2.3$ using slitless spectroscopy from the {\sl HST\/}/WFC3 Infrared Spectroscopic Parallel (WISP) program \citep{atek2010}.  However, even this heterogeneous data set is limited by sample size, as it contains only 91 objects beyond $z \sim 1.85$.
 
The sample of \nsample\ emission-line galaxies presented in \citet{bowman2019}, hereafter referred to as \paperone,
is an excellent database for improving our understanding of \OIII emission at $z \sim 2$ and measuring the epoch's \OIII~$\lambda 5007$ luminosity function.  Since the sample is defined from the \HST/WFC3 G141 grism frames of the 3D-{\sl Hubble Space Telescope} (\threedhst) program \citep{brammer2012, momcheva2016}, the data are similar in both resolution and depth to the planned measurements of the {\sl Euclid} and \RST missions ($R\sim130$, 250, and $550-800$ for \threedhst, {\sl Euclid}, and \RST, respectively).  In addition, the data have been carefully vetted to avoid contamination by low-redshift interlopers and other non-\OIII emitting contaminants.  The sample has comprehensive PSF-matched multi-wavelength photometry \citep{skelton2014}, enabling us to better measure the galaxies' physical properties and the origin of the \OIII emission, and a redshift range ($1.90 < z < 2.35$) that captures both \OII and \OIII emission, thus minimizing selection biases against systems where most of the oxygen is singly ionized.

In this paper, we use the sample defined by \paperone\ to measure the $z\sim2$ \OIII~$\lambda 5007$ luminosity function. In \S\ref{sec:data} we describe the galaxy sample and review how it was defined.  In \S\ref{sec:ew} we use the \threedhst\ line fluxes and the \citet{skelton2014} photometry to measure the galaxies' rest-frame \OIII~$\lambda 5007$ equivalent widths, showing that the distribution is well-fit with an exponential having a scale factor of $\sim 200$\,\AA\null. In \S\ref{sec:lf} we describe our methodology for measuring the sample's  completeness, present the $z \sim 2$ \OIII luminosity function, describe our Bayesian method for fitting the data to an analytical function, and give the best-fitting \citet{schechter1976} parameters.  In \S\ref{sec:number-counts} we use our measurements to estimate galaxy number counts in the $z \sim 2$ universe and in \S\ref{sec:integrated-sfr} we estimate the fraction of the epoch's star formation that is occurring in galaxies with strong \OIII emission. Finally, in \S\ref{sec:discussion}, we summarize the major results and implications of this work.

We adopt a flat-$\Lambda$CDM cosmology with $h=0.7$, $\Omega_M=0.3$, $\Omega_{\Lambda}=0.7$.

\section{Data and galaxy sample}
\label{sec:data}

The data for our analysis consist of a subset of galaxies originally identified by the \threedhst\ survey team \citep{brammer2012, momcheva2016}. The \threedhst\ program combines {\sl HST\/}/WFC3 G141 grism observations ($1.08~\mu{\rm m} < \lambda < 1.68~\mu{\rm m}$) with $0.3~\mu{\rm m} < \lambda < 8.0~\mu{\rm m}$ PSF-matched photometry \citep{skelton2014} over \mbox{$\sim 625$ arcmin$^2$} of sky surveyed by the Cosmic Assembly Near-IR Deep Extragalactic Legacy Survey \citep[CANDELS;][]{grogin2011, koekemoer2011}.  In addition to the grism frames, the 3D-HST's data products include redshift and line-strength catalogs for 79,609 unique sources in AEGIS \citep{davis2007}, COSMOS \citep{scoville2007}, GOODS-N and GOODS-S \citep{giavalisco2004}, and UDS  \citep{lawrence2007}.

\begin{figure*}
  \centering
  \subfloat[]{%
    \includegraphics[height=7.25cm]{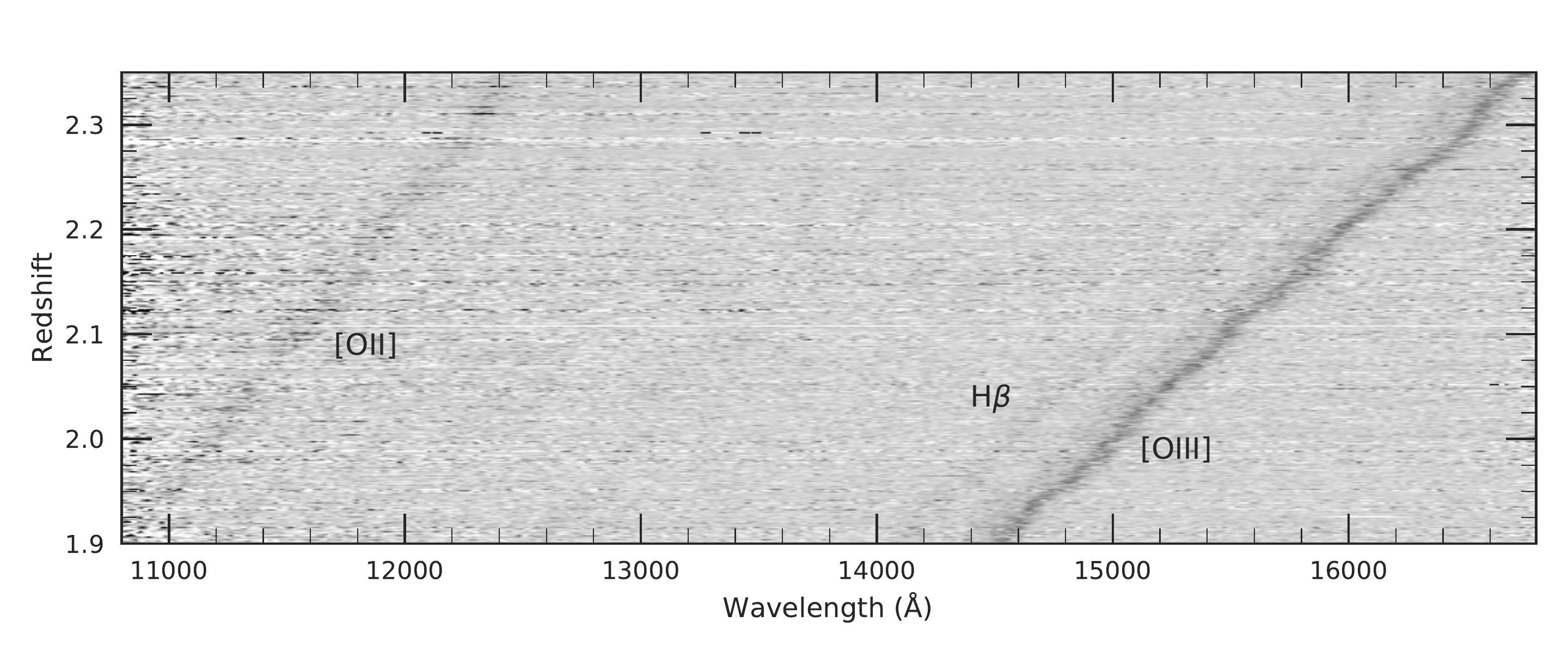}%
  }
  \caption{The \threedhst\ grism spectra for the \nsample\ emission-line selected galaxies presented by \paperone.
  The data have been sorted by redshift, continuum-subtracted via the models provided by \citet{momcheva2016}, and normalized by \OIII flux (which is the brightest feature in $\gtrsim 90\%$ of the galaxies).
  } 
  \label{fig:grism-spectra}
\end{figure*}

\paperone\ vetted this sample to identify emission-line galaxies with IR continuum magnitudes brighter than $m_{\rm J+JH+H} = 26$ \citep[two magnitudes fainter than the vetted catalog of][]{momcheva2016}, no evidence for the presence of an active galactic nucleus (AGN), and unambiguous redshifts based on either (1) the distinctly-shaped asymmetry of the \OIII~$\lambda \lambda 4959,5007$ doublet, (2) two or more obvious emission lines in the 2D grism frames, or (3) one obvious emission line and a well-constrained photometric redshift estimate.   Their resulting data set consists of \nsample\ objects, all with photometric coverage from rest-frame $\sim1200$~\AA\ to $\sim 2.5$~$\mu$m \citep{skelton2014} and grism spectroscopy extending from \OII\ $\lambda 3727$ to \OIII~$\lambda 5007$. In over 90\% of these objects, the blended \OIII~$\lambda\lambda 4959,5007$ doublet is the strongest feature, with the remaining galaxies dominated by either \OII $\lambda 3727$ or \Hb.  In all cases where \OIII\ is not the strongest feature, at least two strong emission lines are clearly present in the spectrum, with over 90\% of the sources having two or more emission lines at a signal-to-noise ratio greater than one. The details of galaxy selection, along with the physical properties of the sample, are given in \paperone\ and \citet{bowman2020}.
Figure~\ref{fig:grism-spectra} displays the grism spectra for the sample.

\section{Equivalent width distribution}
\label{sec:ew}

Emission line surveys are uniquely suited for the identification of star-forming galaxies whose continua are not easily detected in broadband observations.  But their completeness functions can be complicated, as the detection of a galaxy through its line-emission depends on both the flux in the line and the flux density of the continuum. As a result, samples of emission-line galaxies are likely to show incompleteness due to low line-flux, low contrast of the emission line over the continuum (i.e., equivalent width), and, in the case of surveys that involve the pre-selection of targets, low continuum flux density.  The advantage of emission line surveys is that the method is complementary to color selection:  while emission-line searches can identify low-mass systems whose continuum emission is too faint for a photo-$z$ measurement, continuum-selection will catch low-equivalent width objects, sources where dust has extinguished the emission-lines, and galaxies with little or no star formation.

\begin{figure*}
  \centering
  \subfloat[]{%
    \includegraphics[width=0.355\linewidth]{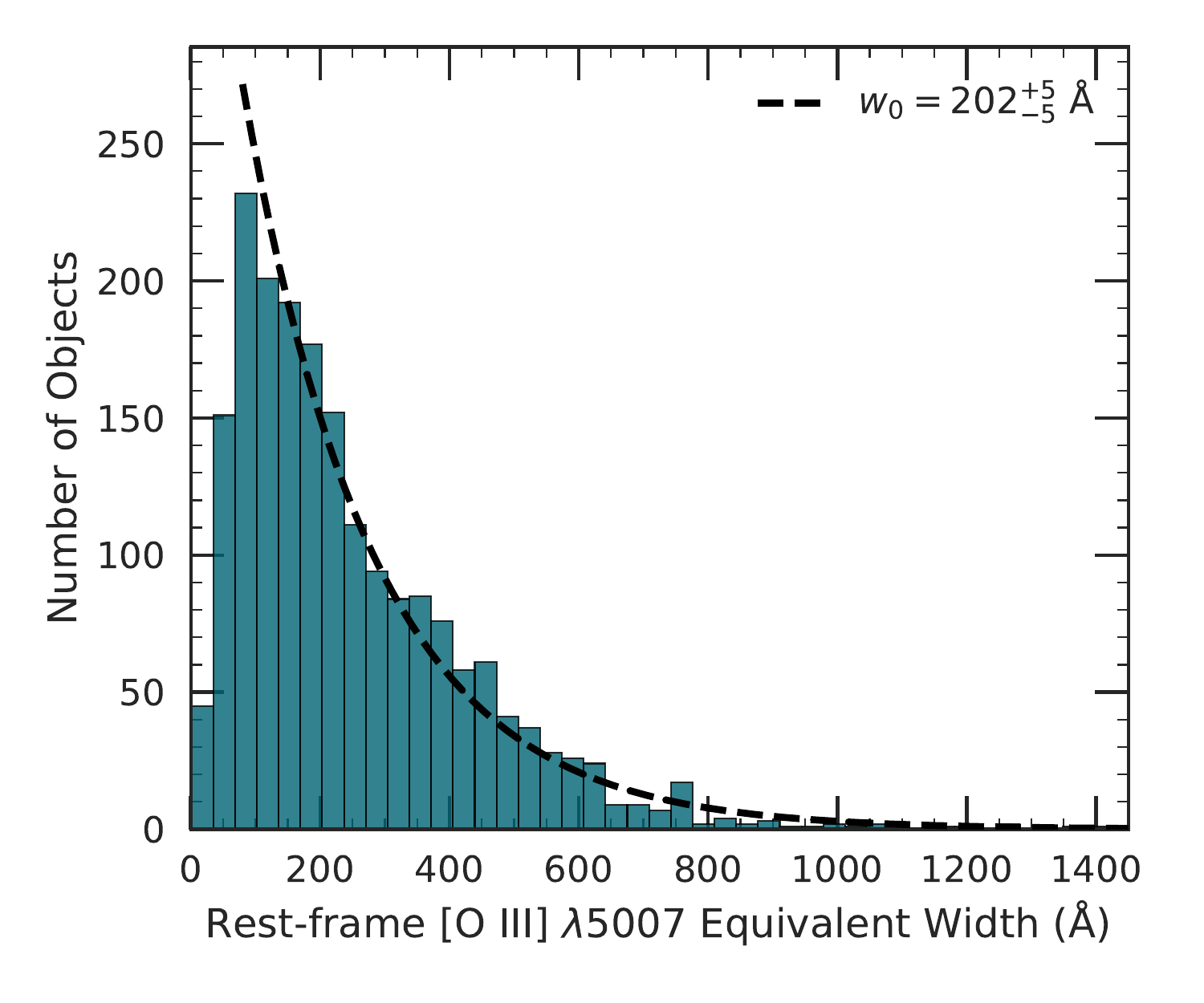}%
  }\hfill
  \subfloat[]{%
    \includegraphics[width=0.645\linewidth]{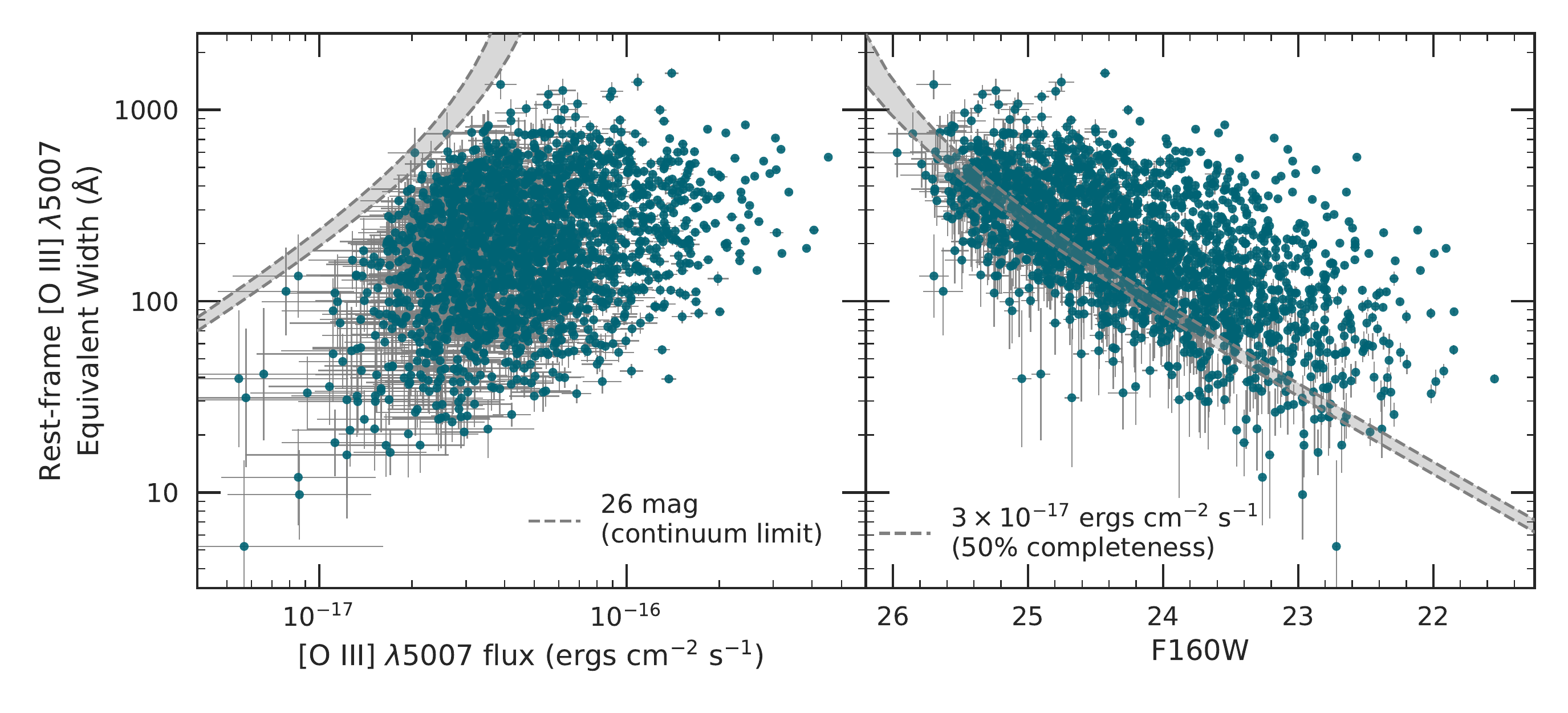}%
  }
  \caption{The left-hand panel displays the distribution of rest-frame \OIII~$\lambda 5007$ equivalent widths, where the continuum is estimated from the {\sl HST\/}/WFC3 F160W flux densities provided by \citet{skelton2014}. The black dashed line shows the exponential fit to the data ($w_0 = 202 \pm 5$\,\AA\null). The center and right-hand panels display the equivalent width measurements as a function of line flux and continuum magnitude, respectively. The grey regions reflect the broadband continuum limit and the 50\% line flux completeness limit across the $1.90 < z < 2.35$ redshift range spanned by the sample.
  }
  \label{fig:oiii-ew}
\end{figure*}

The rest-frame \OIII~$\lambda 5007$ equivalent width distribution of the \paperone\ sample is displayed in the left-hand panel of Figure~\ref{fig:oiii-ew}.  Since the broadband magnitudes of the sample extend to $m_{\rm J+JH+H} = 26$ many of these objects have continua that are too faint to be measured on the 3D-HST grism frames. Thus, the equivalent widths shown in the figure are derived by taking the flux densities inferred from the {\sl HST\/}/WFC3 F160W photometry (which, for our $1.9 < z < 2.35$ galaxies, covers the rest wavelength range between $\sim 4500$ and $\sim 5500$\,\AA), subtracting off the flux contributed by emission-lines within the bandpass, and then comparing this estimated continuum level to the \OIII line strength measured by the G141 grism.  The resultant equivalent width is then divided by 1.33 to account for the contribution of $\lambda 4959$ in the \OIII blend \citep{storey2000}.

The rest-frame equivalent width distribution is well-fit using an exponential with a scale factor of $w_0 \sim 202 \pm 5$\,\AA, and the tail of the distribution extends well beyond $\sim 1000$\,\AA\null. This range of values is similar to that seen by \citet{khostovan2016}, who found rest-frame \mbox{H$\beta$+\OIII} equivalents ranging from $\sim 100$ to 700\,\AA\ in their $z\sim 2.23$ narrowband survey, and \citet{malkan2017}, who measured rest-frame \mbox{\OIII\ $\lambda\lambda 4959,5007$} equivalent widths between $\sim 100$ and 1000\,\AA\ in their $z\sim3$ Lyman-break galaxy sample.

The low-equivalent width regime suffers from incompleteness (displayed in the center and right-hand panels of Figure~\ref{fig:oiii-ew}) for at least two reasons: galaxies with bright continua will have less-obvious emission lines in their grism spectra owing to the higher noise associated with the continuum subtraction, and faint-continua objects may not meet the magnitude cutoff for the sample. Nonetheless, our data set is selected in a manner similar to that expected from the upcoming {\sl Euclid} and \RST surveys, suggesting that the distributions of galaxy properties are likely to be similar.   

\section{\OIII\ Luminosity Function}
\label{sec:lf}

\subsection{Completeness Correction}
\label{sec:lf-completeness}

A common approach for estimating the completeness of a survey is to insert artificial sources onto the data frames, run the same data processing and source selection algorithms as used to construct the true sample, and evaluate the recovery fraction as a function of brightness. However, our sample is identified from the data products of the \threedhst\ survey \citep{brammer2012, momcheva2016}, which have undergone complex data processing and incorporate redshift estimates based upon both the grism observations and the 20 to 40 band photometry of \citep{skelton2014}.  Thus it is not possible to determine completeness from artificial source tests.

Instead, we can estimate the completeness function from the data themselves using common sense constraints. Specifically, we expect that at high line fluxes, where emission-lines are clearly visible, the \threedhst\ sample of galaxies will be essentially complete.  Conversely, at the faint end of the line-flux distribution, emission lines will be easily lost in the noise and the completeness fraction will be low. In between these two extremes, we expect the completeness function to be a well-behaved and monotonic function, without discontinuities or inflection points. 

The function suggested by \citet{fleming1995}
\begin{equation}
F_F(f) = \dfrac{1}{2} \Bigg[1+\dfrac{\xi \log (f/f_{50})}
{\sqrt{1+ [\xi \log (f/f_{50})]^2}}\Bigg]
\end{equation}
satisfies these criteria and is commonly used to model completeness in a wide variety of astronomical surveys \citep[e.g.,][]{salinas2015, lhagen2015, kuzma2016, kim2020}. The function is monotonic, allows full freedom for the steepness of the curve, i.e., it can reproduce a step function if $\xi$ is large, or model a gradual decline in completeness if $\xi$ is small. It is defined by just two parameters:  $\xi$ and $f_{50}$, the latter being the flux at which survey completeness falls to 50\%.

The one minor modification we make to the \citet{fleming1995} curve is at the extreme faint end of the completeness function.  Because the application of the Fleming law to a power law luminosity function severely overpredicts the number of faint objects, we include an additional factor $\tau(f)$, which is defined using $f_{10}$, the location at which the original Fleming function falls to 10\%. This modification depends only upon the two free parameters, $\xi$ and $f_{50}$, so it does not introduce any additional free variables. Thus, our assumed form for the completeness function, $F_C$, is 
\begin{subequations} \label{eq:completeness} 
\begin{align}
F_F(f) &= \dfrac{1}{2} \Bigg[1+\dfrac{\xi \log (f/f_{50})}
{\sqrt{1+ [\xi \log (f/f_{50})]^2}}\Bigg] \label{eq:comp1} \\
\bigskip
\tau(f) &= 1 - e^{-f/f_{10}} \label{eq:comp2} \\
F_C(f) &= \big[F_F(f)\big]^{1 /\tau(f)} \label{eq:comp3}
\end{align}
\end{subequations}
with the two free parameters being $f_{50}$ and $\xi$.  

The above formulae is similar to that used by \paperone\ to estimate completeness in their initial study of $z \sim 2$ grism selected galaxies.  However, while \paperone\ assumed that the underlying flux distribution followed a power law, we release that assumption and jointly model the flux completeness parameters and the luminosity function, as described below. This approach allows us to marginalize over the completeness variables and infer the luminosity function from the data themselves.  However, we note that, in practice, our methodology makes very little difference to the final results:  the luminosity function measured by jointly modeling the completeness curve and the emission-line flux distribution is essentially identical to that which would be obtained by first estimating the flux completeness curve (as in \paperone) and then using those parameters to measure the luminosity function.

\begin{figure}[h!]
\centering
\noindent\includegraphics[width=\linewidth]{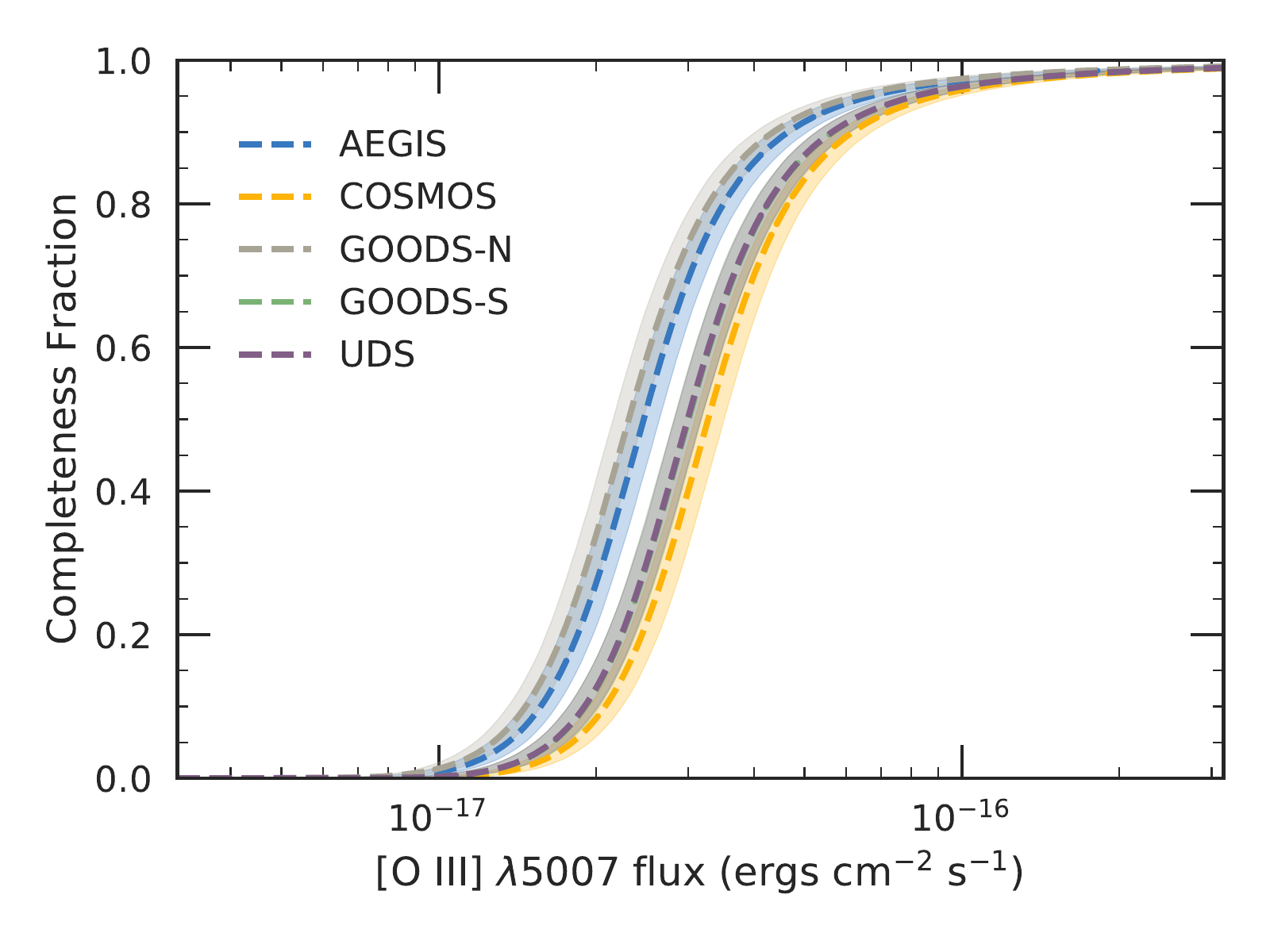}
\caption{The flux completeness curves for each of the five fields. The completeness parameters are estimated by jointly fitting the luminosity function and flux completeness function. The 50\% line-flux completeness limits are \mbox{$\sim 2-3 \times 10^{-17}$~\ecs}\ (the precise values for each field are shown in Figure~\ref{fig:lf-posteriors}).} 
\label{fig:completeness}
\end{figure}

The flux completeness curves that we measure for each field are shown in Figure~\ref{fig:completeness}. This parameterization for the flux completeness function (Equation~\ref{eq:completeness}) has sufficient flexibility to reproduce the completeness curves for other grism surveys, e.g., the Probing Evolution and Reionization Spectroscopically (PEARS) survey (Figure~5 in \citealt{pirzkal2013}). The completeness analysis in \citet{pirzkal2013} used artificial source injection and retrieval simulations and can be well-fit with the parameterization in Equation~\ref{eq:completeness}.

\subsection{Methodology}
\label{sec:lf-methodology}

We use a maximum likelihood estimation (MLE) approach to parameterize the observed \OIII~$\lambda 5007$ luminosity function. 
We use the function first suggested by \citet{schechter1976}, \ie
\begin{equation} \label{eq:LF}
\Phi(L)d(L) = \phi^* (L/L^*)^\alpha e^{-L/L^*} d(L/L^*)
\end{equation}
and follow the MLE prescription described by \citet{ciardullo2013}, which we briefly summarize here.

The observed luminosity function of galaxies is the product of the true luminosity function (Equation~\ref{eq:LF}) and the completeness function (Equation~\ref{eq:completeness}), i.e.,
\begin{equation}\label{eq:LF-observed}
    \Phi'(L,z) = F_C(L,z) \times \Phi(L)
\end{equation}
where $F_C$ has two free parameters ($f_{50}$ and $\xi$) and $\Phi$ is described by $\phi^*$, $L^*$, and $\alpha$.
With this definition, the number of galaxies that we expect to observe in a fixed luminosity and volume interval is $\lambda = \Phi'(L,z) \Delta L \Delta V$.  Due to Poisson statistics, we will actually observe $n$ galaxies, with the likelihood of the observation being  
\begin{equation} \label{eq:LFa}
p(n|\lambda) = \dfrac{\lambda^n e^{-\lambda}}{n!}.
\end{equation}
In the limit of arbitrarily small bins, $n$ will take on the value of either 0 or 1, and the likelihood of observing a given set of objects is
\begin{equation} \label{eq:LFb}
\begin{split}
\mathcal{P} &= \prod_{\substack{\text{bins with}\\\text{zero objects}}} \dfrac{\lambda_i^0 e^{-{\lambda_i}}}{0!} 
    \prod_{\substack{\text{bins with}\\\text{one object}}} \dfrac{\lambda_i^1 e^{-{\lambda_i}}}{1!} \\
  &= \prod_{\text{all bins}} e^{-{\lambda_i}} \prod_i^N \lambda_i
\end{split}
\end{equation}
where $N$ is the total number of observed galaxies in the sample.  When written in terms of differentials, the log likelihood becomes
\begin{equation} \label{eq:LF-ll}
    \ln \mathcal{P} = \sum_i^N \ln \Phi'(L_i,z_i) - \int_{z_{\rm min}}^{z_{\rm max}} \int_{0}^{\infty} \Phi'(L,z) \frac{dV}{dz} dL \, dz
\end{equation}
where $\Phi'(L,z)$ again refers to the observed luminosity function and $dV/dz$ is the comoving volume element.

The most difficult term to compute in the above equation is the volume. Grism surveys are extremely efficient, in the sense that they capture a spectrum for all of the sources within a given pointing. However, this efficiency comes with a burden: unless multiple exposures are taken at different position angles, overlapping spectra will make it impossible to identify the faint emission of a $z \sim 2$ galaxy when it is superposed upon the spectrum of a bright continuum object.  \threedhst 's approach for dealing with the problem was to use the information at hand to model and subtract the spectra of all the overlapping sources.  This procedure helps, but does not eliminate the problem, as even in the best cases, the signal-to-noise of a $z \sim 2$ spectrum will be substantially reduced. 

One solution to this problem is to estimate the fraction of the \threedhst\ survey area where the ability to detect $z \sim 2$ galaxies is compromised.  Using an earlier version of our sample defined by \citet{zeimann2014}, \citet{ciardullo2014} estimated that $\sim 15\%$ of the \threedhst\ footprint is unusable owing to overlapping spectra.  Similarly,  \citet{colbert2013} pointed out that the edges of the \threedhst\ fields are rendered unusable owing to the lack of a direct image or the dispersal of spectra off the detector; they also estimated that the survey area should be reduced by roughly 15\%.

An alternative method for determining the effective area of the \threedhst\ survey is to mask out the regions of each field that could conceivably cause $z \sim 2$ emission-line galaxies to be missed or poorly measured. To do this, we can assume that \textit{every} object in the \threedhst\ catalog ($m_{\rm J+JH+H} \leq 26$) produces a spectrum that renders $z \sim 2$ galaxies unmeasureable.  Clearly, this is an overly-conservative assumption, as the \threedhst\ team's treatment of overlapping spectra can and does allow the detection of emission lines superposed on the continua of nearby objects.  However, this criterion is the cleanest to use: while it does reduce the effective area of the survey by a factor of $\sim 2$, it still leaves us with a sample of \nsampleunmasked\  $z \sim 2$ emission-line galaxies distributed over 345~arcmin$^2$ of sky.

The methodology for creating the grism masks is described in detail by \mbox{Abelson et al.\ (2021, in prep)}.  In short, we create a mask for each G141 grism frame using the list of $m_{\rm J+JH+H} \leq 26$ objects given by \citet{skelton2014} as a source catalog.  We first determine the sky position and orientation of each 3D-HST frame, as well as the position of each source appearing within its frame.  We then model the expected positions of these objects' spectra as rounded rectangles, with the rectangle's width equal to the dispersion length and its height dependent on the brightness of the source. A pixel is considered masked if it falls within one of these rounded rectangles; by extension, a source is considered masked if its spectrum falls on top of masked pixels.  If a source is present in more than one pointing, it is only counted as masked if it is masked in every pointing, since the object will be detectable in the pointing where its spectrum is not overlapped by that of another source.

We note that because galaxies are inherently clustered, the self-masking of objects at a common redshift can, in theory, alter the true galaxy luminosity function (and, more importantly, the measurement of the galaxy correlation length).  However, this effect is not important for our analysis, as only a small fraction of associated objects will be located along the direction of spectral dispersion. In fact, given the low density of $1.90 < z < 2.35$ emission-line galaxies ($\sim 3$~arcmin$^2$) on the grism frames, the self-masking of one galaxy in this redshift range by another is quite rare.  The effect should therefore have minimal impact on the parameters of the luminosity function.

\subsection{Results}
\label{sec:lf-results}

We fit the \OIII data of the five CANDELS fields simultaneously. We assume that the steepness of the completeness curve, $\xi$, is the same across all five fields and allow for a different 50\% line flux limit ($f_{50}$) for each field.  These assumptions result in a nine-parameter model: three parameters that describe the luminosity function, one that describes the steepness of the completeness curve, and five values of the 50\% flux completeness limits (one for each field).

As discussed in \S\ref{sec:lf-methodology}, the normalization of the luminosity function from grism surveys is nontrivial. Thus, we first adopt the approach used by other authors and include our entire sample of \nsamplefull\  objects in our analysis, but we reduce the effective area of the survey by $\sim 15\%$ \citep[e.g.,][]{colbert2013, ciardullo2014}. Using this assumption, we find best-fit values of $\log L^* = 42.50 \pm 0.06$, $\log \phi^* = -2.69 \pm 0.10$, and $\alpha = -1.51 \pm 0.15$.  Alternatively, if we apply the mask described in Section~\ref{sec:lf-methodology}, our galaxy sample and volume are both reduced by roughly one-half, but the end results are consistent: the faint-end slope steepens by $\sim 0.1$ and $L^*$ brightens by $0.06$~dex, both of which are statistically consistent with the results from the full sample. The volume density of sources, derived by integrating each solution down to $L = 10^{41}$~\ecs, decreases by $\lesssim 20\%$ and suggests that the 15\% area reduction may be mildly overestimated. In what follows, we use the results based on the \nsampleunmasked\ objects that remain after applying the mask.

\begin{figure*}[hp!]
\centering
\noindent\includegraphics[width=\linewidth]{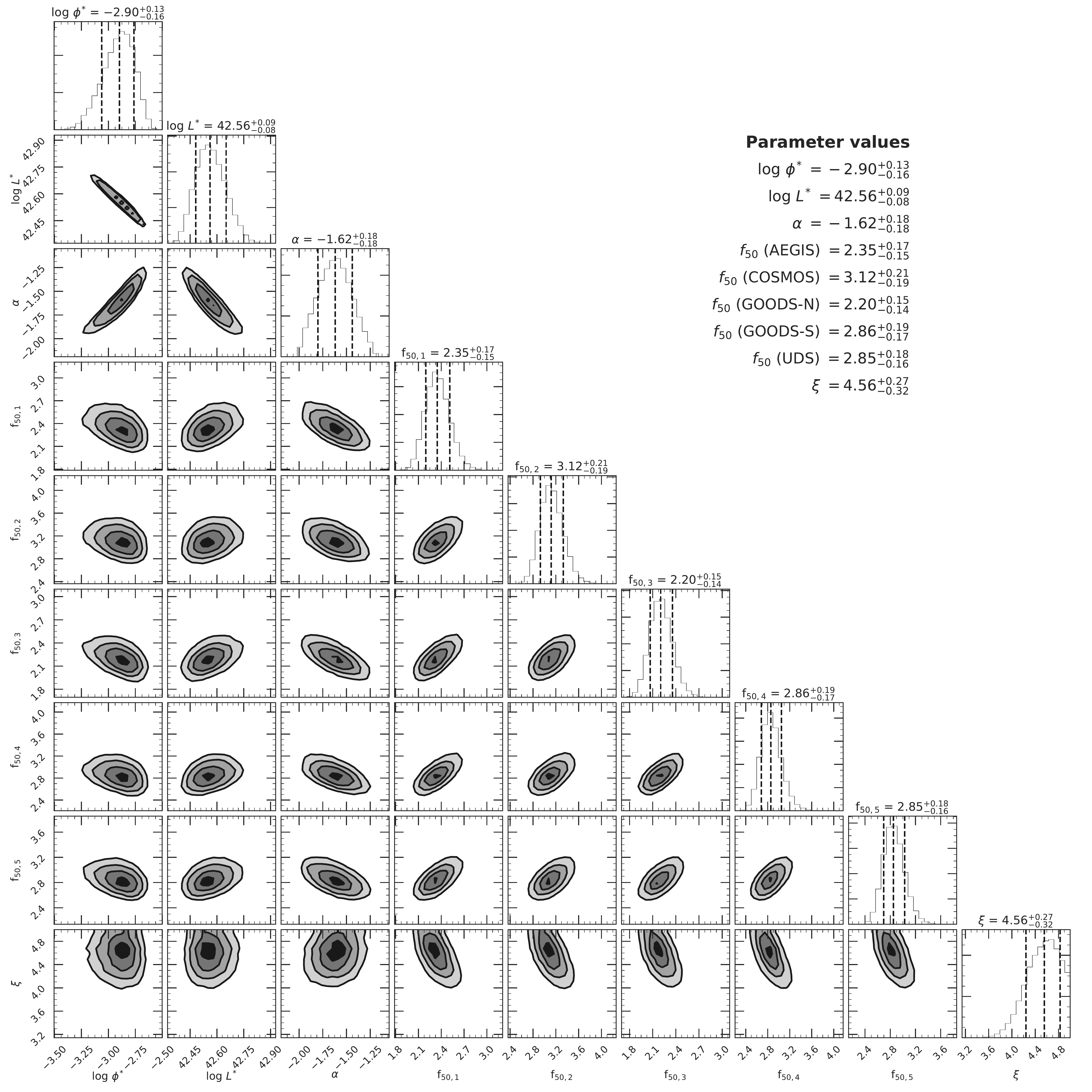}
\caption{The joint posterior distribution for the \citet{schechter1976} parameters ($\phi^*$, $L^*$, and $\alpha$) and the flux completeness curve parameters ($f_{50}$ and $\xi$). The $f_{50,i}$ estimates correspond (in order) to the 50\% flux limits for the AEGIS, COSMOS, GOODS-N, GOODS-S, and UDS fields, and are reported in units of $10^{-17}$~\ecs. The outermost contours show the $2\,\sigma$ levels and the contours move inward in steps of $\sigma/2$. The top panel in each column displays the marginalized posterior distribution of each parameter, with the dashed lines depict the 16\%, 50\%, and 84\% quantiles. The \citet{schechter1976} parameters $L^*$ and $\phi^*$ are largely decoupled from the completeness parameters, while the faint-end slope shows a weak degeneracy. The 50\% flux limits and the steepness of the completeness curve have opposing effects on $\alpha$, so marginalizing over the completeness parameters yields a stable solution for the \citet{schechter1976} fit. Estimates of the \citet{schechter1976} parameters are unchanged if we remove sources below these 50\% flux limits and refit the model to the remaining objects, indicating that completeness corrections to faint sources are not driving the solution.} 
\label{fig:lf-posteriors}
\end{figure*}

Figure~\ref{fig:lf-posteriors} displays the joint posterior distributions of our fit. The panels indicate that the three luminosity function parameters, while showing strong degeneracies among themselves, are largely decoupled from the variables that affect completeness. Since $\alpha$ and the completeness parameters both depend on the faint end of the distribution, there is a weak degeneracy between them; however, $f_{50}$ and $\xi$ have opposing effects on $\alpha$, so marginalizing over the completeness curve yields a stable result for this faint-end slope. We do note, however, that the faint end of the galaxy luminosity function is notoriously difficult to constrain, given its dependence on the completeness correction and the strong degeneracy with the other two \citet{schechter1976} parameters. This difficulty is reflected in the range of values given in the literature: at $z < 1$, estimates for $\alpha$ range from $\alpha \sim -1.22 \pm 0.13$ \citep{ly2007} to $\alpha \sim -1.83 \pm 0.1$ \citep{comparat2016}, while at $z \sim 2$ the limited data available suggest $\alpha \sim -1.6$ \citep{colbert2013, mehta2015}, albeit with large uncertainties on the estimates. Our measurement of $\alpha$ should be treated with a degree of caution, as the \OIII~$\lambda 5007$ 50\% flux limit ($\sim 3 \times 10^{-17}$~\ecs) is only $\sim 0.5$~dex fainter than the best-fit value of $L^*$; nonetheless, the agreement with other $z\sim2$ estimates \citep{colbert2013, mehta2015} is encouraging.

It is standard practice in luminosity function studies to exclude faint sources, i.e., those requiring large completeness corrections. To ensure that the low-completeness regime is not driving our estimates of the \citet{schechter1976} parameters, we also perform a fit with all objects lying below the 50\% flux completeness limits removed.  The results from the remaining \nsamplefluxcutfitmasked\ objects are completely consistent with those using the full sample, with each of the \citet{schechter1976} parameters consistent to within $0.5\,\sigma$ of the previous estimates. This consistency indicates that our best-fit luminosity function is not driven by our choice for the form of the faint-end of the completeness function.

\begin{figure}[h!]
\centering
\noindent\includegraphics[width=\linewidth]{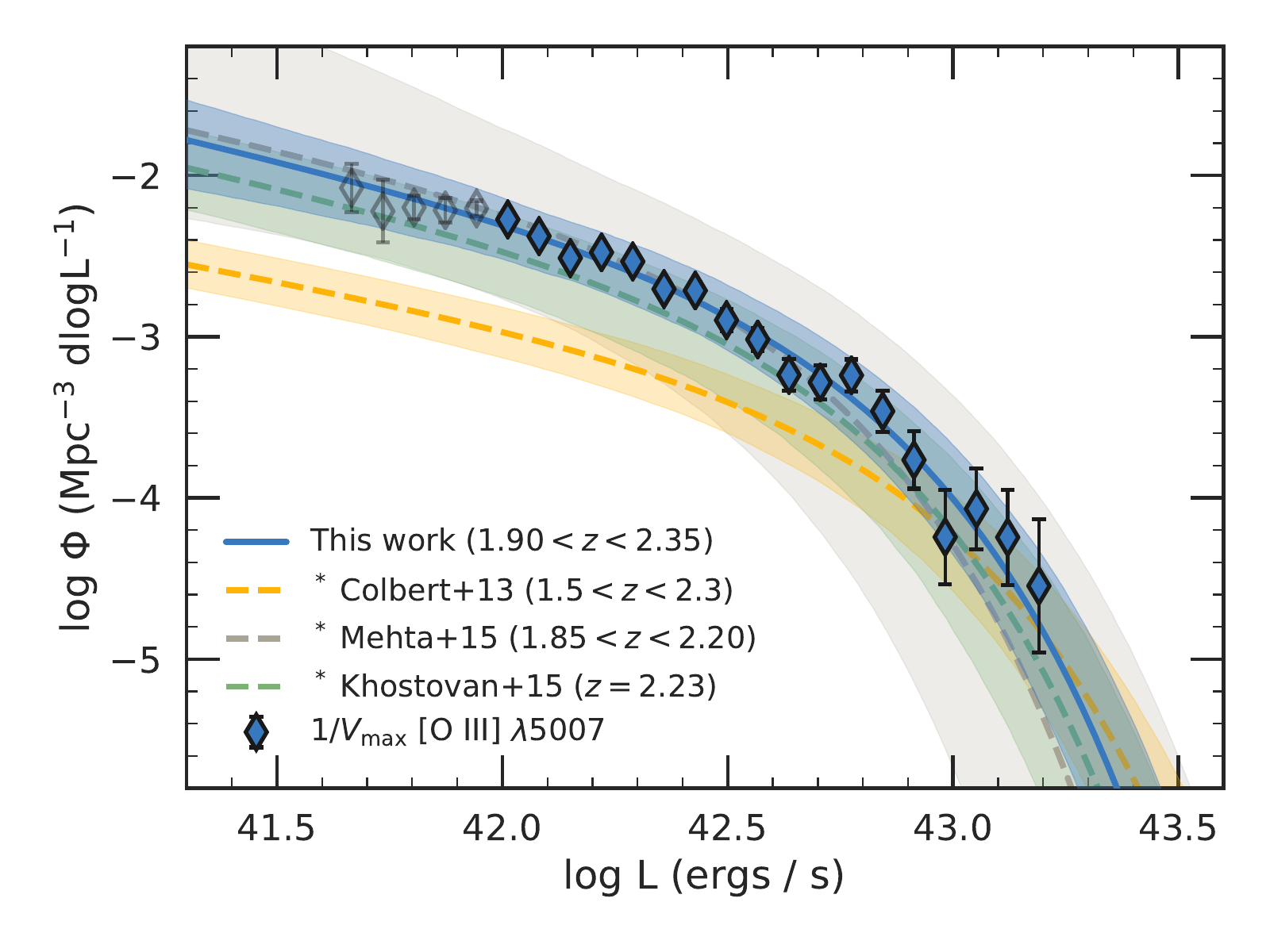}
\caption{The \OIII~$\lambda 5007$ luminosity function for our $z\sim2$ sample of galaxies. The most likely \citet{schechter1976} fit to the \nsampleunmasked\ objects after applying the mask is shown as the blue curve, while the shaded region reflects the $1\,\sigma$ uncertainty realized by drawing \citet{schechter1976} parameters from their posterior distributions. The data points are realized using the 1/$V_{\rm max}$ estimator, while our best-fit curve is computed using the procedure described in \S\ref{sec:lf-methodology}. The transparent points consist primarily of objects that fall below the 50\% line flux completeness limits. Also shown are the fits from other $z \sim 2$ studies; in each case, the literature curves have been scaled to reflect only the emission of the \OIII~$\lambda 5007$ emission line.  The bright end of our function agrees with the results from \citet{mehta2015} and \citet{khostovan2015}, while the luminosity function of \citet{colbert2013} has a value of $L^*$ that is $\sim 0.2$~dex brighter than that of the other studies. Our estimate of $\alpha$ is in excellent agreement with \citet{colbert2013} and \citet{mehta2015}, but should be treated with a degree of caution since the 50\% flux limit of our sample is only $\sim 0.5$~dex fainter than the best-fit value of $L^*$.
} 
\label{fig:lf-best}
\end{figure}

As noted above, the error contours in Figure~\ref{fig:lf-posteriors} imply that the variables that describe the \citet{fleming1995} completeness curve are largely decoupled from those associated with the \citet{schechter1976} function.  We can therefore use those variables to measure the shape of the $z \sim 2$ \OIII luminosity function nonparametrically using the $1/V_{\rm max}$ technique \citep{schmidt1968, huchra1973}. In this measurement, the data points at the bright end are unaffected by our completeness estimates; the precise values for completeness only effect the faint end of the distribution. These data points are displayed in Figure~\ref{fig:lf-best}, where we compared 1/$V_{\rm max}$ binned data to our best-fit \citet{schechter1976} solution and other \OIII\ luminosity functions that appear in the literature. Table~\ref{tab:LF-params} compiles our best-fit \citet{schechter1976} parameters. In order to compare luminosity functions, we reduced the literature values of $L^*$ by $\sim 0.126$~dex to account for contamination of \OIII~$\lambda 5007$ by its companion line at 4959\,\AA\null.  While this correction is strictly true for the grism measurements of \citet{colbert2013} and \citet{mehta2015}, who quoted the combined flux of the \OIII doublet \citep{storey2000}, it is only approximately correct for the narrowband measurements of \citet{khostovan2015}, which include both \OIII and H$\beta$. (The authors state that the contribution from \Hb\ is negligible at the bright end of the luminosity function.)

Figure~\ref{fig:lf-best} demonstrates that our assumption of a \citet{schechter1976} parameterization for the $z \sim 2$ \OIII\ luminosity function is a good one. Despite the limited sample sizes of the previous measurements, the systematics of object selection, and the effects of cosmic variance, all luminosity functions shown in Figure~\ref{fig:lf-best} are in good agreement.  In particular our luminosity function is an excellent match to the functions derived by \citet{mehta2015} and \citet{khostovan2015}, and our value of $L^*$ is only moderately lower ($\sim 0.2$~dex) than that found by \citet{colbert2013}.  

One effect that may cause the luminosity functions to disagree is the inclusion of AGN:  while deep X-ray images of the CANDELS fields allowed \paperone\ to remove most AGN from their sample, such data do not exist for the WISP observations \citep{colbert2013, mehta2015}. Moreover, although \citet{khostovan2015} targeted two of the regions observed by CANDELS, those fields (COSMOS and UDS) have the shallowest X-ray data of the survey.  (The authors considered other indicators of AGN activity, such as rest-frame near-IR colors and the distribution of AGN luminosities, but this only serves to highlight the differences between various measurements of bright-end of the luminosity function.)  In our case, if the \nagnfull\ AGN identified by \paperone\ are re-inserted into the sample, the value of $L^*$ increases by only $\sim 0.1$~dex while giving a similar-quality fit. 

A more important effect may be cosmic variance.  By performing spectroscopic follow-up on the galaxies found in a 10~deg$^2$ narrowband survey, \citet{sobral2015} empirically characterized the uncertainties in the \citet{schechter1976} parameters arising from cosmic variance.  For the space volume of our survey, they estimate that the uncertainties can be as large as $\sim 20\%$ for $L^*$ and $\sim 40\%$ for $\phi^*$ at $z=2.2$. However, as discussed in detail in \citet{colbert2013}, slitless spectroscopic surveys such as \threedhst\ have some advantages when it comes to the issue of cosmic variance. Since grism observations survey beam-like windows (i.e., a wide redshift range within a relatively small sky area), as opposed to the more box-shaped volumes characteristic of narrowband studies, the objects are generally spread over a larger range of environments owing to the longer line of sight \citep{trenti2008}. Further helping the situation is that our sample lies in five distinct, uncorrelated fields, which effectively reduces the error contribution from cosmological variance by a factor of $\sqrt{5}$ \citep{colbert2013}. For our sample, the uncertainty arising from cosmic variance is estimated to be $\lesssim 15\%$ \citep{trenti2008}.

\begin{figure}[h!]
\centering
\noindent\includegraphics[width=\linewidth]{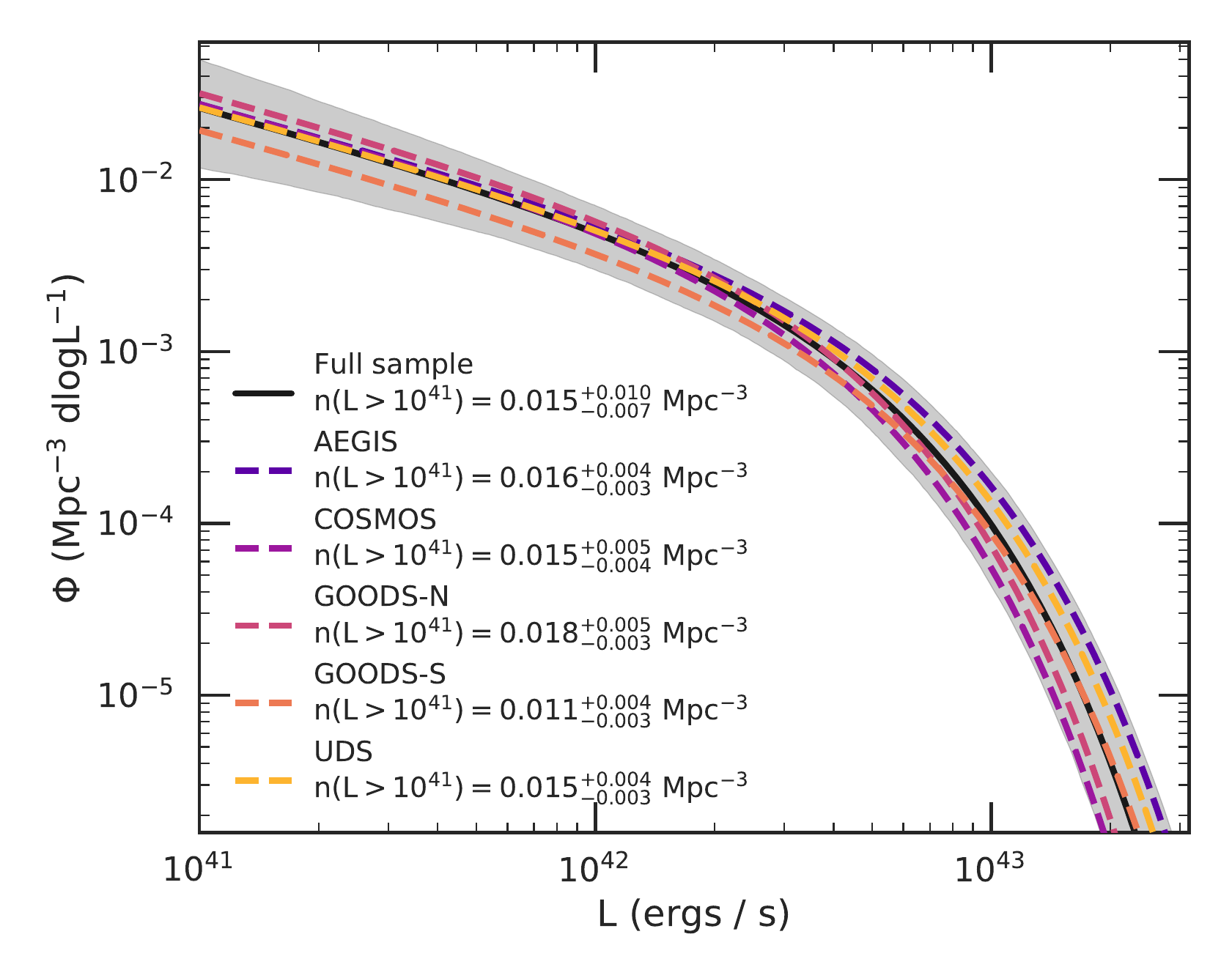}
\caption{The best-fit \OIII~$\lambda 5007$ luminosity functions measured for the full set of objects lying outside of the grism mask (black solid curve and grey shaded region) and for each of the five fields individually (dashed lines). The flux completeness parameters and faint-end slope, measured from fitting all five fields together (Figure~\ref{fig:lf-posteriors}), are fixed when fitting the luminosity function in the individual fields, and only the objects above the 50\% flux limits are used in the fit. The best-fit $L^*$ values vary by 0.17~dex (from $\log L^* = 42.47$ to $42.64$). The expected number densities, estimated by integrating the luminosity functions down to $L=10^{41}$~\ecs, are displayed in the legend and vary by $\sim 20\%$ relative to the fit to all five fields.
} 
\label{fig:lf-cv}
\end{figure}

Alternatively, we can make our own determination of cosmic variance by using the fact that our $z \sim 2$ data set is derived from observations in five different areas of the sky.  For this test, we assign the completeness parameters and faint-end slope that were measured from the joint fit to all five fields (Figure~\ref{fig:lf-posteriors} and Table~\ref{tab:LF-params}), fit for the \citet{schechter1976} parameters in each field individually, and compare each field's best-fit solution.  The results are displayed in Figure~\ref{fig:lf-cv}, with the legend indicating the expected number density of sources derived from integrating the luminosity function down to $10^{41}$~\ecs.  The best-fit $L*$ values vary by 0.17~dex (from $\log L^* = 42.47$ to $42.64$) and the best-fit number densities of \OIII\ emitters range from 0.011 to 0.018~galaxies~Mpc$^{-3}$.  This $\sim 20\%$ scatter around the full sample normalization of 0.015~galaxies~Mpc$^{-3}$ is consistent with expectations.  

Observational uncertainties can also affect the estimated shape of the luminosity function, particularly at the bright end. \citet{mehta2015} address this point by introducing a modified maximum likelihood estimator that utilizes the full probability distribution of each luminosity measurement, as opposed to treating each as a point estimate. We performed a similar test to ensure that our estimate of the luminosity function is robust against observational uncertainties. We simulated \nsimsample\ random samples by drawing line fluxes from a Gaussian distribution centered at each measured value and having a standard deviation equal to the recorded error.  We then refit the model (\S\ref{sec:lf-methodology}) for each simulated set of fluxes in the same manner as the actual data.  

\begin{figure}[h!]
\centering
\noindent\includegraphics[width=\linewidth]{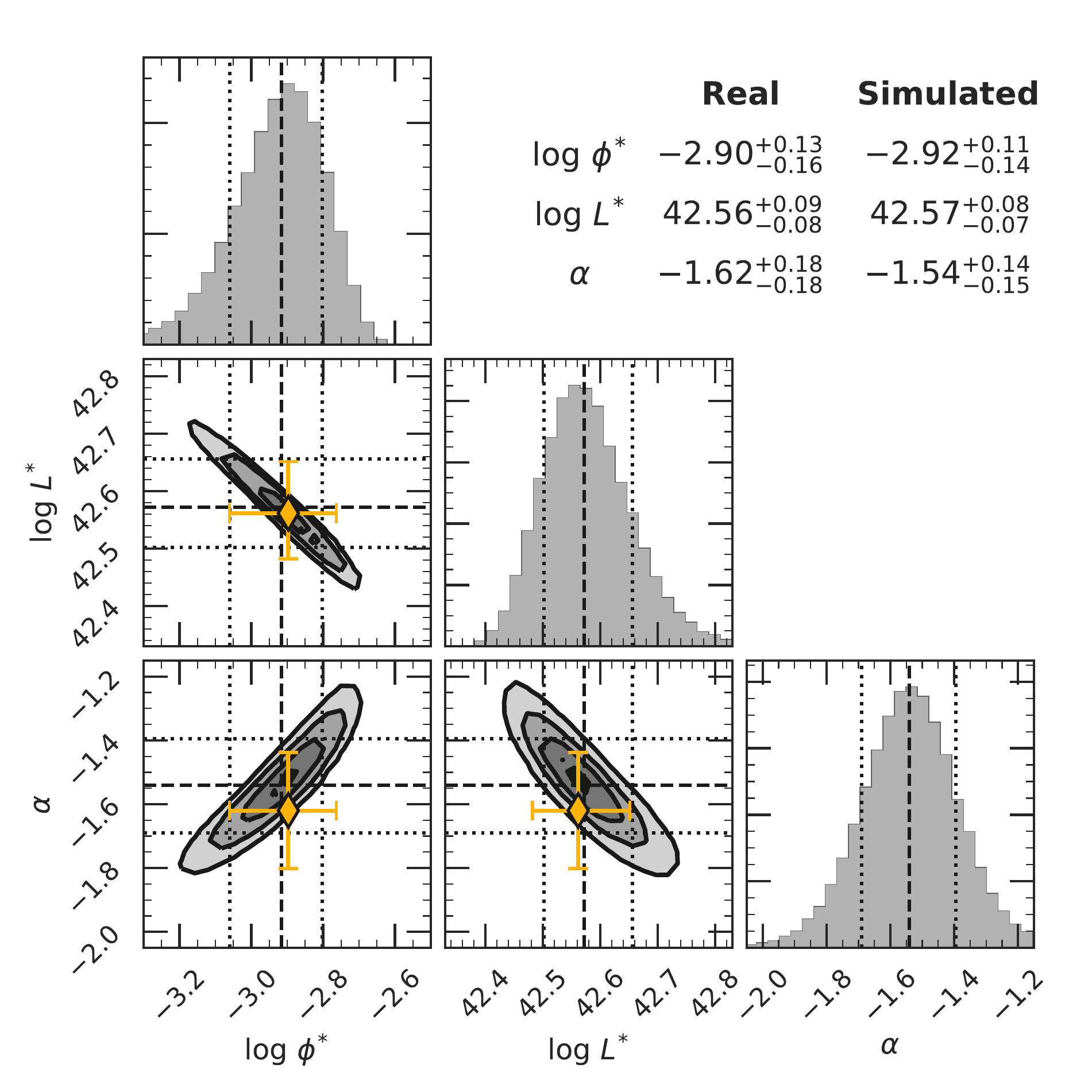}
\caption{The joint posterior distributions of $\phi^*$, $L^*$, and $\alpha$ for \nsimsample\ simulated data sets, each constructed by perturbing the original emission-line fluxes within their reported errors and remeasuring the luminosity function. The median values and $1\,\sigma$ uncertainties for the simulated sets are indicated by the dashed and dotted black lines, while the amber diamonds display the MLE solution from the observed fluxes. The estimates for the \citet{schechter1976} parameters from the original and simulated data are in good agreement, suggesting that measurement errors are not skewing our parameter estimates.
} 
\label{fig:lf-error-sim}
\end{figure}

Figure~\ref{fig:lf-error-sim} displays the joint posterior distributions for all of the simulated samples, with the median values and $1\,\sigma$ uncertainties shown via the dashed and dotted lines, respectively.  Also shown are the values (and uncertainties) of each \citet{schechter1976} parameter measured from the unperturbed data set.  The figure demonstrates that measurement errors are not corrupting our estimates of the luminosity function parameters.

\begin{deluxetable*}{lcccccc}
\tablecaption{The maximum likelihood \citet{schechter1976} parameters for the \OIII\ luminosity function of our sample, as well as relevant comparisons from the literature. We fit the five fields simultaneously and jointly fit the luminosity function and flux completeness curves, as described in \S\ref{sec:lf-methodology}. When fitting the dust-corrected luminosity function, we fix the completeness parameters and faint-end slope measured from the set of observed fluxes after applying the mask, and discard observations below the 50\% flux completeness limits. The results for our sample refer to only the \OIII~$\lambda 5007$ line. 
\label{tab:LF-params}}
\tablehead{
\colhead{Reference} & \colhead{Redshift} & \colhead{Sample size} &  \colhead{$\log \phi^*$} & \colhead{$\log L^*$} & \colhead{$\alpha$} & \colhead{Sample notes}
\\
\colhead{} & \colhead{} & \colhead{} &  \colhead{(Mpc$^{-3}$)} & \colhead{(\ecs)} & \colhead{} & \colhead{}
}

\startdata
\citet{colbert2013} & $1.5 - 2.3$ & 58 & $-3.74 \pm 0.43$ & $42.91 \pm 0.37$ & $-1.67 \pm 0.78$ & WISPS, \OIII~$\lambda\lambda 4959,5007$ \\
\citet{colbert2013} & $1.5 - 2.3$ & 58 & $-3.60 \pm 0.14$ & $42.83 \pm 0.11$ & $-1.5$ (fixed) & WISPS, \OIII~$\lambda\lambda 4959,5007$ \\
\citet{mehta2015} & $1.85 - 2.20$ & 91 & $-2.69_{-0.51}^{+0.31}$ & $42.55_{-0.19}^{+0.28}$ & $-1.57_{-0.77}^{+0.28}$ & WISPS, \OIII~$\lambda\lambda 4959,5007$ \\ 
\citet{khostovan2015} & 2.23 & 271 & $-3.03_{-0.26}^{+0.21}$ & $42.66_{-0.13}^{+0.13}$ & $-1.60$ (fixed) & HiZELS, \Hb\ + \OIII \\
\hline
 This study (observed) & $1.90 - 2.35$ & \nsamplefull\ & $-2.69_{-0.11}^{+0.09}$ & $42.50_{-0.06}^{+0.06}$ & $-1.51_{-0.16}^{+0.15}$ & 85\% area \\
 This study (observed) & $1.90 - 2.35$ & \nsampleunmasked\ & $-2.90_{-0.16}^{+0.13}$ & $42.56_{-0.08}^{+0.09}$ & $-1.62_{-0.18}^{+0.18}$ & Mask applied \\
 This study (dust-corrected) & $1.90 - 2.35$ & \nsamplefluxcutfitmasked\ & $-3.69_{-0.05}^{+0.05}$ & $43.38_{-0.05}^{+0.06}$ & $-1.62$ (fixed) & Mask applied \\
\enddata
\end{deluxetable*}

\section{Number counts}\label{sec:number-counts}

One of the most important applications of a luminosity function lies in predicting the number of galaxies that future cosmology surveys like {\sl Euclid} and \RST\ will find.  Since the achievable precision of these surveys depends on the number of galaxies identified (and the bias of the tracers, squared), it is useful to cast our luminosity function measurement in these terms.  For consistency with  previous measurements, we estimate the number of galaxies that will be detected (per square degree) across the redshift range of our sample ($1.90 < z < 2.35$).

The {\sl Euclid} Wide Survey (WS) will find \OIII\ emitting galaxies between $1.5 \lesssim z \lesssim 2.7$ down to a flux limit of \mbox{$\sim 2 \times 10^{-16}$~\ecs}.
In contrast, {\sl Euclid}'s Deep Survey (DS) and the \RST\ High-Latitude Spectroscopic (HLS) survey will both have completeness limits comparable to the 50\% line flux limit of our \threedhst\ sample, with estimated depths of $f_{\rm lim} \sim 6 \times 10^{-17}$~\ecs\ \citep{vavrek2016, spergel2015}.  Because these limits straddle the location of $L^*$ in our redshift range, estimating the number density of $1.90 \lesssim z \lesssim 2.35$ galaxies that {\sl Euclid} and \RST will detect is particularly sensitive to the shape of the bright end of the luminosity function.
 
\begin{figure}[h!]
\centering
\noindent\includegraphics[width=\linewidth]{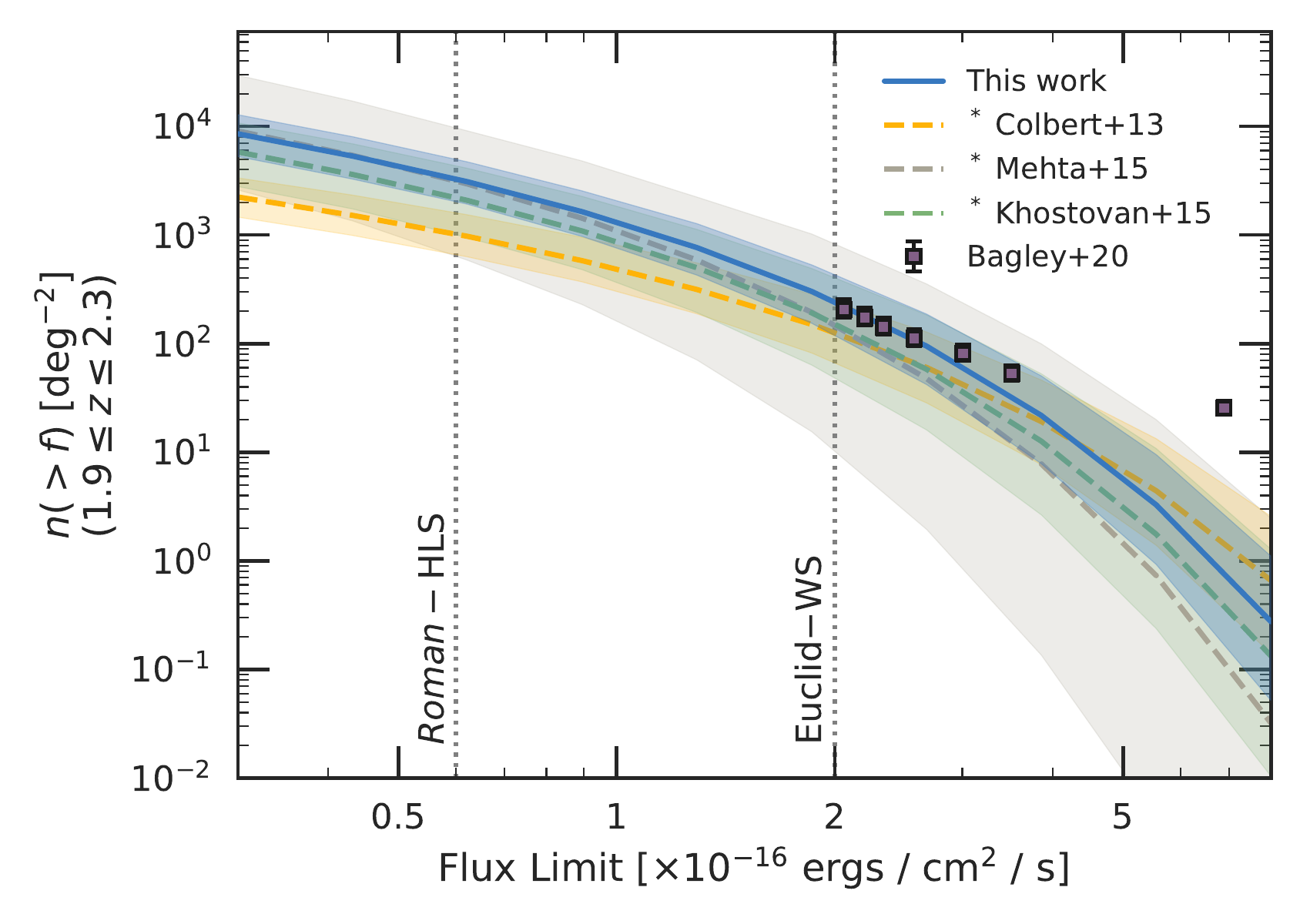}
\caption{The surface density (deg$^{-2}$) of $1.9 \leq z \leq 2.3$ \OIII~$\lambda 5007$ emitters as a function of flux limit. The data points from \citet{bagley2020} are based directly on the number counts for their sample (which is culled from the WISPS, \threedhst , and AGHAST grism surveys), while the remaining curves and $1\,\sigma$ shaded regions are realized using the \citet{schechter1976} luminosity function parameters. We remove the contribution of the \OIII~$\lambda 4959$ line from the $L^*$ estimates in the literature \citep{colbert2013, mehta2015, khostovan2015} by reducing their values of $L^*$ by $0.126$~dex. The expected flux limits of the {\sl Euclid}-WS and \RST -HLS programs are illustrated by the grey dotted lines.
} 
\label{fig:surface-density-5007}
\end{figure}

Figure~\ref{fig:surface-density-5007} shows the surface density of galaxies between $1.9 \leq z \leq 2.3$ as a function of the limiting flux. With the exception of \citet{bagley2020}, who directly use the number counts from an emission-line sample drawn from both the WISPS and \mbox{\threedhst+AGHAST} surveys, these curves (and confidence intervals) are realized using the \citet{schechter1976} parameters (and corresponding uncertainties) presented in the referenced papers. The curves refer to the ``true'' number, i.e., they directly adopt the \citet{schechter1976} parameters presented in the other studies, which are estimated from completeness-corrected data. As above, we correct the luminosity functions from the previous studies for the contribution of the \OIII~$\lambda 4959$ line by reducing the reported $L^*$ values by 0.126~dex; while this correction is not exact for the H$\beta +$\OIII\ luminosity function from \citet{khostovan2015}, the contribution of H$\beta$ to galaxies at the bright end of the \OIII luminosity function is expected to be minimal.

All five of the estimates plotted in Figure~\ref{fig:surface-density-5007} are in relatively good agreement at the flux limit of the {\sl Euclid} Wide Survey and predict a surface density of $1.9 \leq z \leq 2.3$ \OIII emitters of between 150 and 250 galaxies~deg$^{-2}$.  Our result and that of \citet{bagley2020} predict slightly higher values than the other three analyses. At brighter and fainter fluxes, the estimates diverge, with the former discrepancy driven by sample size and survey volume, and the latter caused by difficulties associated with estimating the faint-end slope. Below $\sim 2 \times 10^{-16}$~\ecs, our result is in good agreement with the predictions from \citet{mehta2015} and \citet{khostovan2015} (i.e., within a factor of $\lesssim 2$), while \citet{colbert2013} predicts 3 to 5 times fewer galaxies at $6 \times 10^{-17}$~\ecs, due to their shallower (fixed) faint-end slope. Since none of the previous $z\sim2$ \OIII\ luminosity functions have been able to tightly constrain the faint end of the luminosity function, this regime remains uncertain. 

\section{Star Formation Rate Density}\label{sec:integrated-sfr}

The global star formation rate density (SFRD) of the Universe encodes information about the manner in which galaxies form and evolve. Measurements of this quantity across cosmic time indicate that the SFRD peaked between $2 \lesssim z \lesssim 3$ (see the review by \citet{madau2014} and references therein), although the systematics associated with the various star formation rate (SFR) indicators and the different treatments of dust complicate the picture. While \OIII~$\lambda 5007$ is a somewhat controversial tracer of SFR in the local Universe \citep[e.g.,][]{moustakas2006}, there is reason to be more optimistic at higher redshifts \citep{kaasinen2018}.  

Independent of redshift, the recombination lines of hydrogen are an excellent tracer of the ionizing photons coming from hot young stars.  In contrast, while the forbidden transitions of metals such as oxygen and nitrogen depend on the SFR, their strengths are also affected by  conditions in the interstellar medium, the ionization state of the gas, and the abundance of the element in question.  Moreover, in the local universe, most of the oxygen in \ion{H}{2} regions is singly ionized, meaning that while \OII is an adequate tracer of star formation, even a small change in conditions can greatly affect the strength of \OIII~$\lambda 5007$.  In the high-redshift universe, the opposite is true, as O$^{++}$ is the dominant species \citep[see, for example the review by][and references therein]{kewley2019}. 

\subsection{Dust Correction}

In order to estimate the intrinsic volumetric star formation rate of our sample, we first have to correct the observed \OIII luminosities for the effect of dust. The assumptions about the behavior of dust attenuation systematically impact the inferences made about galaxies, yet the properties of attenuation remain one of the most important outstanding questions in extragalactic astronomy, especially at high redshift. The effect of dust on the emergent flux of a galaxy depends sensitively on several factors (e.g., dust geometries, dust grain sizes, viewing angles, column densities) and can vary both systematically, as a function of the physical conditions within the galaxy, and stochastically across different galaxies. To further complicate the matter, dust impacts the different components of a galaxy (i.e., young/old stars, nebular gas) in different ways \citep[e.g.,][]{calzetti2000, charlotfall2000}. To address the problem, a number of attenuation laws have been proposed in the literature \citep[e.g.,][]{cardelli1989, calzetti2000, battisti2016, reddy2020}, with the main differences between the laws being the greyness of the curves, the amount of excess attenuation around 2175\,\AA, and the relationship between the attenuation that affects starlight and that which applies to emission lines (see \citealt{shivaei2020} for a complete discussion).

\begin{figure}[h!]
\centering
\noindent\includegraphics[width=\linewidth]{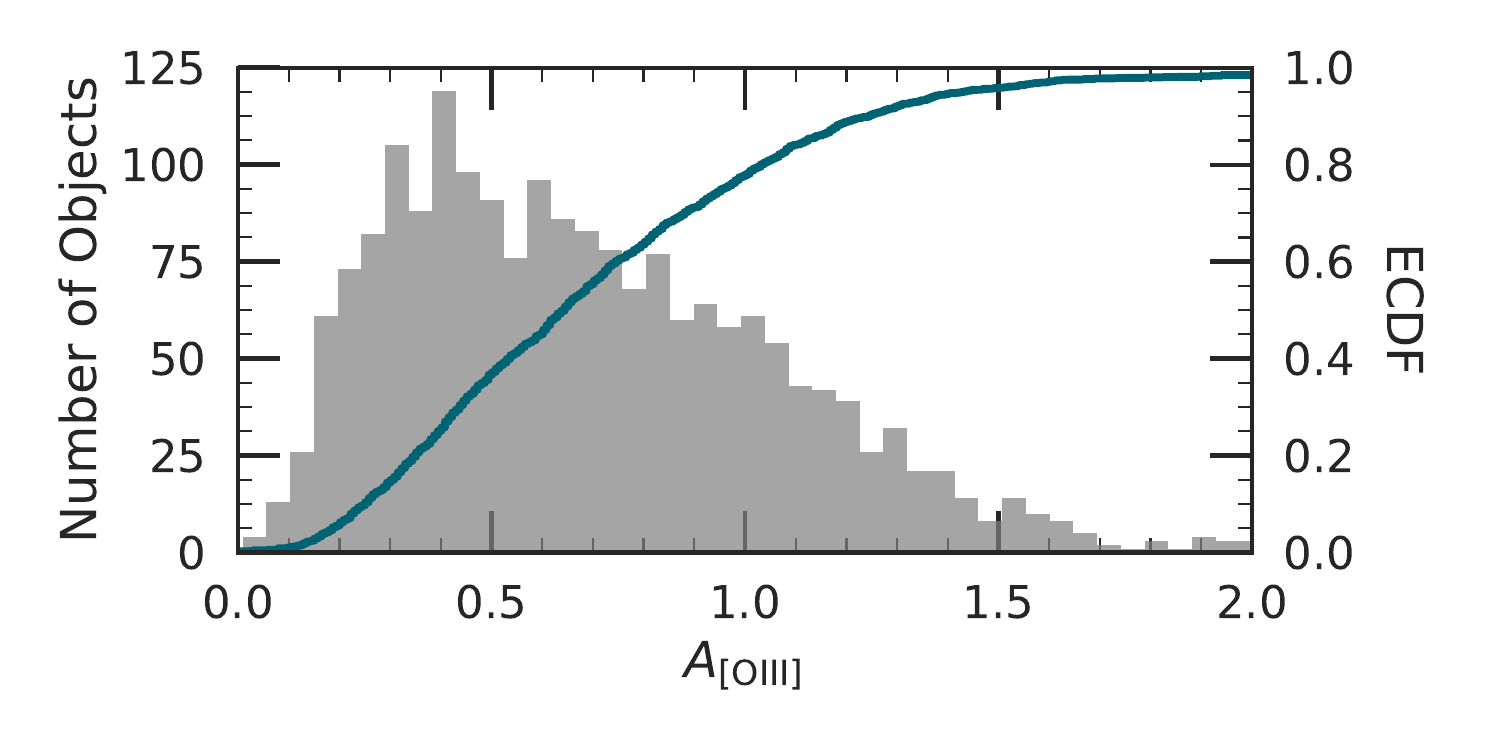}
\caption{ The distribution of $A_{\rm [OIII]}$ values for the galaxies in our sample. The histogram shows the differential distribution, while the solid cyan curve gives the cumulative fraction.  The line extinctions are estimated from the stellar reddening estimates (derived from the objects' spectral energy distributions; \citealt{bowman2020}) via the
\citet{reddy2020} nebular attenuation law.} 
\label{fig:a-oiii}
\end{figure}

We deredden the observed \OIII~$\lambda 5007$ luminosities using the nebular attenuation law from \citet{reddy2020}, who studied a sample of $\sim 500$ $1.6 \lesssim z \lesssim 2.6$ star-forming galaxies identified in the MOSFIRE Deep Evolution Field survey. We use the relation in \citet{reddy2020} to convert the stellar reddening estimates derived from each galaxy's spectral energy distribution (see \citealt{bowman2020} for details) into nebular reddening, and then into $A_{\rm [O\,III]}$ via
\begin{equation} \label{eq:ebv_a5007}
\begin{split}
E(B-V)_{\rm nebular} &= 2.07 \times E(B-V)_{\rm stellar} \\
A_{\rm [O\,III]} &= 3.46 \times E(B-V)_{\rm nebular} 
\end{split}
\end{equation}
The distribution of $A_{\rm [O\,III]}$ values is displayed in Figure~\ref{fig:a-oiii}.  The figure illustrates that the median \OIII\ attenuation is close to $\sim 0.5$~mag, although there is a tail that extends to attenuations as great as $\sim 2$~mag.

\subsection{Calibrating the SFR Relation}

\begin{figure}[h!]
\centering
\noindent\includegraphics[width=\linewidth]{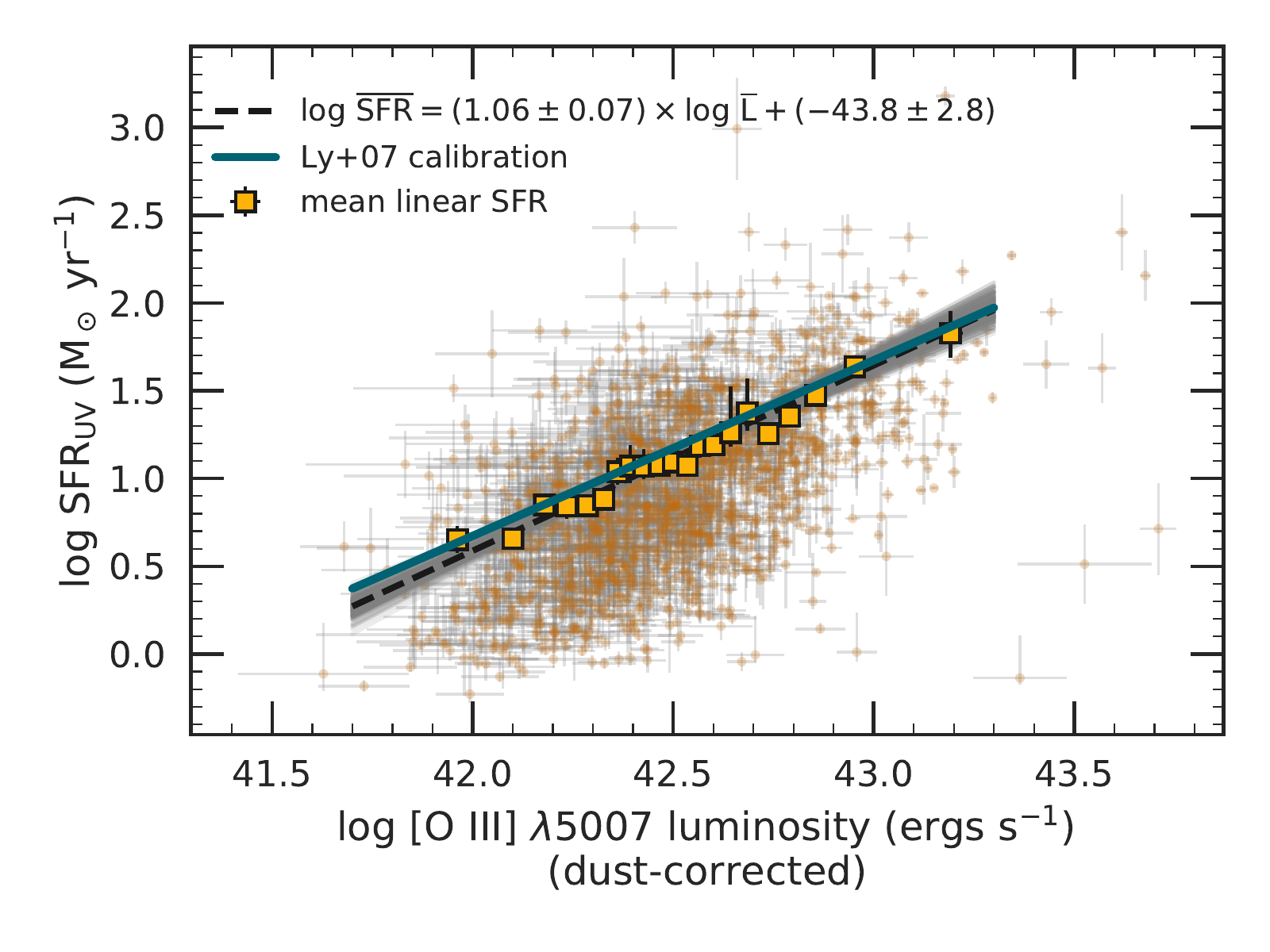}
\caption{ The relationship between our dust-corrected \OIII\ luminosities and UV SFRs.  
The amber points show the mean linear SFR in each luminosity bin.  These are used to derive our relation between dust-corrected \OIII\ luminosity and UV SFR (depicted by the black dashed curve). The \citet{ly2007} calibration is included for reference (cyan line) and has been corrected to a \citet{kroupa2001} initial mass function.
} 
\label{fig:sfr-LOIII}
\end{figure}

The extensive photometric data that are available for our sample enable measurements of the SFR for each object, allowing us to calibrate the relation between dereddened \OIII luminosity and SFR in the $z \sim 2$ universe.  To do this, we use the SFRs found by \paperone\ from the galaxies' rest-frame ultraviolet (UV) flux densities, UV spectral slopes, and the SFR calibration of \citet{kennicuttandevans2012}, i.e.,
\begin{equation}
    \log {\rm SFR}_{\rm UV} = \log L_{1600} -43.35\ M_{\odot}\ {\rm yr}^{-1}
\end{equation}
where $L_{1600}$ is the dust-corrected 1600~\AA\ luminosity density.  We then compare these SFRs to the galaxies' dereddened \OIII luminosities.  Figure~\ref{fig:sfr-LOIII} shows the results of this comparison.  

As Figure~\ref{fig:sfr-LOIII} shows, a $z \sim 2$ galaxy's \OIII\ luminosity, when corrected for extinction following Equation~\ref{eq:ebv_a5007}, is well-correlated with its SFR, albeit with a dispersion in the log of $\sim 0.4$~dex at fixed luminosity. Since a log-log space fit (i.e., a minimization using logarithmic scatter) will underestimate the mean SFR at a given luminosity, we derive a linear relation between the two quantities. We do this by binning the data by \OIII  luminosity, computing the mean linear SFR within each bin, and fitting a linear relation to these binned data.  The result is
\begin{equation} \label{eq:sfr-L-calib}
\log {\rm SFR_{UV}} = (1.06\pm0.07) \times (\log L_{\rm [O\,III], 0}) - (43.8 \pm 2.8)
\end{equation}
where $L_{\rm [O\,III], 0}$ is the dust-corrected \OIII~$\lambda 5007$ luminosity. The uncertainties in the slope and intercept are estimated via 5000 bootstrap iterations, with each iteration determined by measuring the ordinary least squares bisector \citep{feigelson1992}. In other words, we bootstrap the original sample, bin by luminosity (as described above), and measure the slope and intercept on each resulting set of binned data. Our calibration is consistent with others that appear in the literature, e.g., \citet{ly2007}.

\subsection{Computing the SFRD}
We can use the relationship between \OIII luminosity and SFR (Equation~\ref{eq:sfr-L-calib}), and the dust-corrected \citet{schechter1976} parameters (Table~\ref{tab:LF-params}) to estimate how much of the epoch's star formation is occurring in \OIII emitting galaxies.  There are two ways to do this calculation.
 
First, the data in Figure~\ref{fig:a-oiii} can be used to correct the \OIII luminosity of each galaxy individually, enabling a measurement of the dust-corrected \OIII luminosity function. The difficulty with this direct approach is that it introduces a systematic error associated with completeness, as the fraction of galaxies missing from a SFR-limited sample is magnitude-dependent.  (In other words, at faint magnitudes, a larger fraction of galaxies are attenuated out of the sample.)  While fixing the faint-end slope can partially mitigate the problem, the issue still propagates into the integrated SFR via the degeneracy between $L^*$ and $\alpha$.

A common alternative for calculating SFRD is to apply the same extinction correction to every galaxy in the sample, often $A_{\rm H\alpha}=1$ \citep[e.g.,][]{hopkins2004, ly2007, matthee2017}. The distribution of $A_{\rm [OIII]}$ for our sample peaks near 0.5~mag (Figure~\ref{fig:a-oiii}) and since $A_{\rm [OIII]} / A_{\rm H\alpha} = 1.3$ for the \citet{reddy2020} law, the median attenuation of the galaxies in this study is lower than $A_{\rm H\alpha}=1$. If we use the stellar reddening values to estimate $A_{\rm H\alpha}$ and take the approach outlined above, we find that a uniform $A_{\rm H\alpha}=1$ would overestimate the dust correction for $\sim 92\%$ of our objects and therefore overestimate the epoch's integrated star formation rate. For this reason, and because we have dust estimates for individual galaxies, we adopt the first approach of dereddening each galaxy separately using the relations given in Equation~\ref{eq:ebv_a5007}.

We fit the dust-corrected \OIII\ luminosities of the \nsamplefluxcutfitmasked\ objects in unmasked regions that lie above the 50\% flux completeness limits to a \citet{schechter1976} function and give the results in  Table~\ref{tab:LF-params}.  Since we do not have any way of deriving the completeness correction versus luminosity for our dust-corrected measurements, we fix the faint-end slope to $\alpha=-1.62$, which is the best-fit value found by our maximum-likelihood analysis of the observed luminosity distribution.  Given the nontrivial effect of dust content on luminosity completeness, we do not attempt to fit for $\alpha$ using the dust-corrected luminosities.

We use Equation~\ref{eq:sfr-L-calib} and the best-fit parameters of Table~\ref{tab:LF-params} to estimate the UV-based SFRD within the \OIII emitting galaxies of the $1.90 < z < 2.35$ universe.  If $\Phi(L)$ is the \citet{schechter1976} function of Equation~\ref{eq:LF}, then 
\begin{equation} 
\rho_{\rm SFR} = \int_{0}^{\infty}\ \Phi(L)\ {\rm SFR_{UV}}(L)\ dL
\label{eq:L-density}
\end{equation}
where ${\rm SFR_{UV}}$ is the relation given in Equation~\ref{eq:sfr-L-calib}.

\begin{figure}[h!]
\centering
\noindent\includegraphics[width=\linewidth]{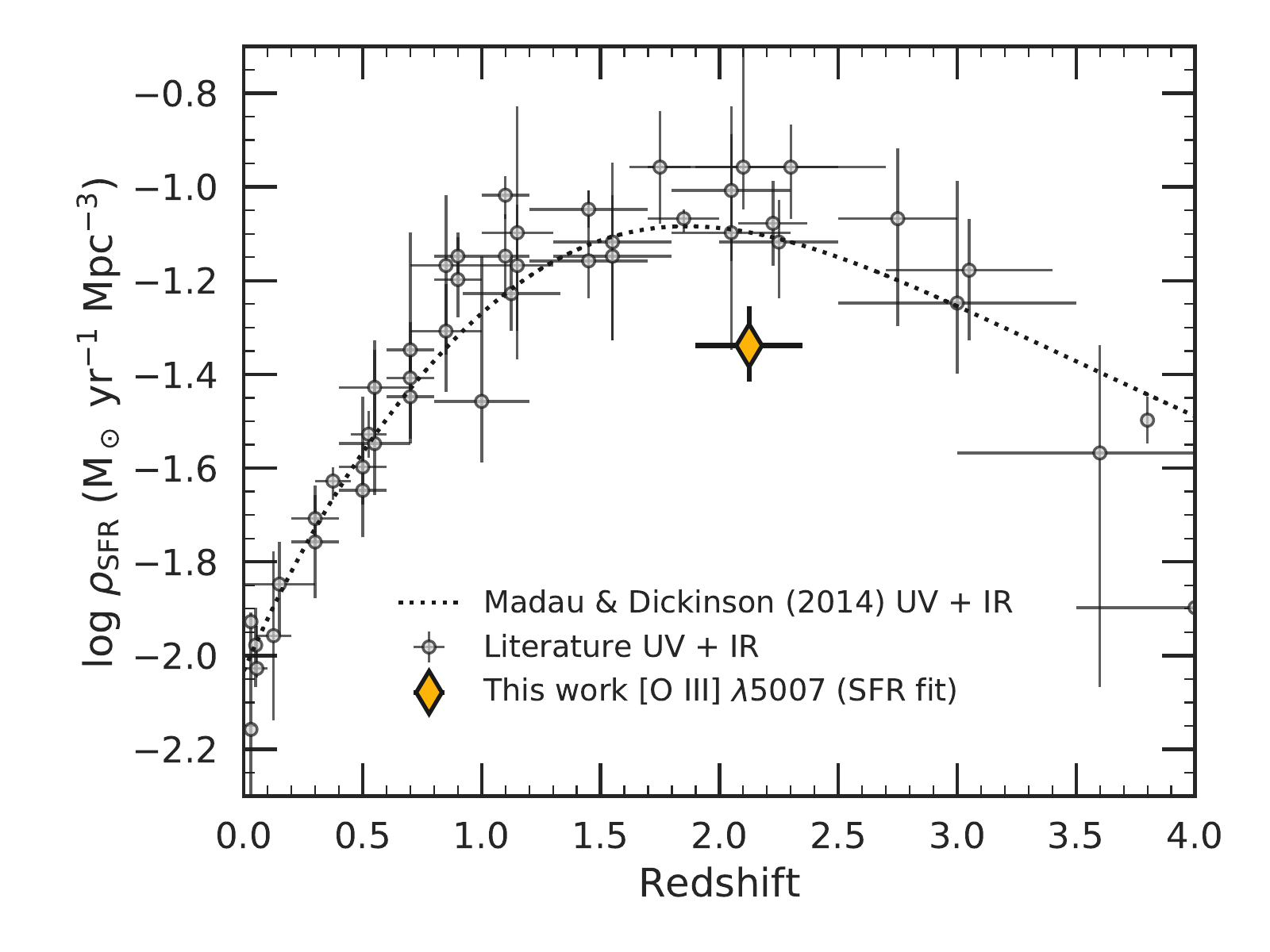}
\caption{The star formation rate density for $z \sim 2$ \OIII-emitting galaxies (amber diamond) compared to global SFRD of the universe. The \OIII value is computed using our calibration between attenuation-corrected \OIII luminosity and UV star formation rate (Equation~\ref{eq:sfr-L-calib}) and adopts the best-fit \citet{schechter1976} function given in Table~\ref{tab:LF-params} with the faint-end slope fixed to $\alpha=-1.62$. The grey points show the UV+IR measurements compiled by \citet{madau2014} and the black dotted curve shows their parameterization, corrected to assume a \citet{kroupa2001} initial mass function. Our measurement is $\sim 0.25$~dex below the UV+IR value, suggesting that \OIII emitting galaxies contain roughly half of the epoch's total star-formation activity.
} 
\label{fig:sfr-density}
\end{figure}

Figure~\ref{fig:sfr-density} compares our SFRD to literature measurements compiled by \citet{madau2014}.  Our estimate of \mbox{$\log \rho_{\rm SFR} = -1.33\pm0.08$ \Msun\ yr$^{-1}$ Mpc$^{-3}$} with $\alpha=-1.62$ is $\sim 0.25$~dex below their UV+IR calibration, suggesting that \OIII galaxies between $1.90 < z < 2.35$ contain $\sim 60\%$ of the epoch's total (UV+IR) star formation activity. This fraction is consistent with the results of \paperone, who found that their emission-line sample only contained $\sim 30\%$ of the galaxies with photometric redshifts between $1.90 < z < 2.35$ in the \threedhst\ catalog. While a subset of the photometric redshift sample is quiescent, the data of \citet{sherman2020} show that the quiescent fraction is not sufficient to explain the discrepancy. The \paperone\ analysis also found strong trends with rest-frame near-IR continuum magnitude and dust content, where the fraction of emission-line galaxies fell to zero at bright magnitudes and large $A_V$ values. Dust is likely a driving factor leading to this lower estimate of the \OIII SFRD compared to the value found from UV+IR studies.

The assumed value of the faint-end slope is a source of significant systematic error in this calculation, and since estimating the completeness correction (as a function of dust and luminosity) is nontrivial, we have no alternative but to fix the value of $\alpha$. As demonstrated by the parameters in Table~\ref{tab:LF-params} and the $\alpha$-$L^*$ contour displayed in Figure~\ref{fig:lf-posteriors}, a steeper faint-end slope results in a brighter estimate of $L^*$ and, thus, a higher SFRD\null. Measurements of the UV luminosity function are subject to this same source of uncertainty, as most studies are not deep enough to tightly constrain the faint-end slope of the galaxy luminosity function. Many studies find (or fix) the value near $\alpha=-1.60$; for example, \citet{alavi2014} used a sample of gravitationally lensed galaxies at $z\sim 2$ and measured a faint-end slope of $\alpha = -1.56 \pm 0.13$. Clearly, the precise fraction of the total star formation activity occurring in an \OIII-selected sample is influenced by poorly-constrained systematics of the calculation.

\section{Conclusion}
\label{sec:discussion}

We have measured the \OIII~$\lambda 5007$ luminosity function for a vetted sample of \nsamplefull\ $1.90 < z < 2.35$ galaxies found in the \threedhst\ grism survey and selected on the basis of their strong rest-frame optical emission lines, with \OIII\ being the strongest feature in $\gtrsim 90\%$ of the sample. Since the volume correction for grism surveys is difficult to estimate, we also apply a grism mask and fit the remaining \nsampleunmasked\ sources. We jointly measure the luminosity function and flux completeness parameters and find best-fit values of $\log \phi^* = -2.90\pm0.15$, $\log L^* = 42.56 \pm 0.09$, and $\alpha = -1.62 \pm 0.18$. However, we caution that the sample's 50\% flux completeness limit extends only $\gtrsim 0.5$~dex past the epoch's observed value of $L^*$, allowing for only moderate constraints on the faint-end slope.

We use our luminosity function to estimate the surface density of \OIII sources in the $1.9 < z < 2.3$ universe. Our values are in good agreement with other estimates in the literature \citep{colbert2013, mehta2015, khostovan2015, bagley2020}. We estimate that in the redshift range $1.9 < z < 2.3$, there are $\sim 200$ galaxies per square degree brighter than $2\times 10^{-16}$~\ecs\ and $\sim 5000$ galaxies brighter than $6\times 10^{-17}$~\ecs.   These are the nominal flux limits of the {\sl Euclid}-WS  and \RST-HLS programs, respectively \citep{vavrek2016, spergel2015}. 

We have presented a measurement of the star formation rate density associated with $z \sim 2$ galaxies that emit in \OIII $\lambda 5007$. Our \OIII-SFR calibration, which is based upon a comparison between the galaxies' attenuation-corrected UV luminosity densities and their dereddened \OIII luminosities, is consistent with other calibrations that appear in the literature. Our calibration between dust-corrected \OIII~$\lambda5007$ luminosity and UV SFR is reasonably well-constrained, but translating that relation into an estimate of $\rho_{\rm SFR}$ is sensitive to assumptions about the faint end of the luminosity function.  Our best-fit measurement of $\alpha = -1.62$ is in agreement with the values found by \citet{colbert2013} and \citet{mehta2015}, and implies an \OIII SFRD that is only $\sim 60\%$ that which is derived from the rest-frame UV+IR\null.  

In the coming era of large grism spectroscopic surveys like {\sl Euclid} and \RST, we will find millions of $z>1$ galaxies via the strength of their rest-frame optical emission lines. The interpretation of these data for galaxy evolution studies will require accurate measurements of their line luminosity functions.  The \OIII luminosity function measured here directly informs the observing strategy that is required to meet the primary science goals of these missions, namely, constraining the nature of dark energy via the measurement of the galaxy power spectrum.  

\acknowledgments

We thank Joel Leja for insightful discussions regarding these results and Micaela Bagley for sharing their galaxy number counts. This work was supported by the NSF through grant AST-1615526 and through NASA Astrophysics Data Analysis grant NNX16AF33G, and was based on observations taken by the CANDELS Multi-Cycle Treasury Program with the NASA/ESA HST, which is operated by the Association of Universities for Research in Astronomy, Inc., under NASA contract NAS5-26555.  The data were obtained from the Hubble Legacy Archive, which is a collaboration between the Space Telescope Science Institute (STScI/NASA), the Space Telescope European Coordinating Facility (STECF/ESA), and the Canadian Astronomy Data Centre (CADC/NRC/CSA). The Institute for Gravitation and the Cosmos is supported by the Eberly College of Science and the Office of the Senior Vice President for Research at the Pennsylvania State University.

\facility{HST (WFC3)}

\software{Astropy \citep{astropy:2018}, NumPy \citep{numpy}, SciPy \citep{scipy}, Matplotlib \citep{matplotlib}, emcee \citep{emcee}}

\bibliographystyle{aasjournal}
\bibliography{elg}

\end{document}